\documentclass[12pt]{article}
\usepackage{booktabs}
\usepackage{pdflscape}
\usepackage{rotating}
\usepackage{caption}
\usepackage{fancyhdr}
\usepackage{longtable}
\usepackage{tabu}
\usepackage{rotating} %needed for sideways table
\usepackage[section]{placeins}   %

\usepackage[yyyymmdd,hhmmss]{datetime}
\usepackage[pagebackref]{hyperref}  %RMT uncomment start here for hyperref
\hypersetup{ colorlinks,
    citecolor=blue,
    filecolor=blue,
    linkcolor=blue,
    urlcolor=blue
}
\renewcommand*{\backref}[1]{}

\newcommand{\captionl}[1]{\captionsetup{width=.9\textwidth}{\caption[dummy]{{\sf #1}}}}

\usepackage{xcolor}

\definecolor{gray}{rgb}{0.6,0.6,0.6}
\definecolor{red}{rgb}{0.85,0,0}
\definecolor{green}{rgb}{0,0.85,0}
\definecolor{blue}{rgb}{0,0,0.85}
\definecolor{beige}{rgb}{0.92,0.87,0.78}
\usepackage[all]{hypcap}    %causes link to figures to go to figure, not caption
\usepackage{overcite}       %med phys superscript like references
\usepackage{graphicx}
\graphicspath{ {./} }

\makeatletter \renewcommand\@biblabel[1]{$^{#1}$} \makeatother
 \newlength{\bibhang}
\newcommand{\mfig}[1]{\marginpar{{\sf Fig~\ref{#1} }}}
\newcommand{\mtab}[1]{\marginpar{{\sf Table~\ref{#1} }}}
\newcommand{\cen}[1]{\begin{center} #1 \end{center}}

\usepackage[mathlines]{lineno}
\newcommand{\ie}{{\it i.e.}, }
\newcommand{\eg}{{\it e.g.}, }

\newcommand{\io}{{${}^{125}$I }}
\newcommand{\pa}{{${}^{103}$Pd }}
\newcommand{\ioc}{{${}^{125}$I}}
\newcommand{\pac}{{${}^{103}$Pd}}

\newcommand{\cs}{{${}^{131}$Cs }}
\newcommand{\cscc}{{${}^{131}$Cs}}

\newcommand{\pd}{\pa}

\newcommand{\eb}{{\tt egs\_brachy }}
\newcommand{\ebc}{{\tt egs\_brachy}}
\newcommand{\bd}{{\tt BrachyDose }}
\newcommand{\bdc}{{\tt BrachyDose}}
\newcommand{\g}{{\tt g }}

\newcommand{\BD}{{\tt BrachyDose }}

\newcommand{\MCNP}{{\tt MCNP}}
\newcommand{\mcnp}{{\tt MCNP }}
\newcommand{\mcnpc}{{\tt MCNP}}

\newcommand{\E}{{CLRP\_EPv2 }}
\newcommand{\e}{{CLRP\_EPv1 }}

\newcommand{\ec}{{CLRP\_EPv1}}

\newcommand{\db}{{CLRP\_EPv2 }}
\newcommand{\dbc}{{CLRP\_EPv2}}

\begin{document}

%Title page

\cen{{\Large {\bfseries  Update of the CLRP eye plaque brachytherapy database for photon-emitting sources}} \\
\vspace{1mm}

\pagestyle{empty}
\pagenumbering{roman}
\vspace*{1mm}
Habib Safigholi, Zack Parsons, Stephen G. Deering, and Rowan~M.~Thomson \\
Carleton Laboratory for Radiotherapy Physics, Department of Physics,\\ 
Carleton University, Ottawa, Ontario, K1S 5B6, Canada\\
\vspace{2mm}

email: a) safigholi@gmail.com    ~ and  ~ b) rthomson@physics.carleton.ca 
 
%\mbox{~}\hfill Last latexed \today\ at \currenttime 
%\vspace{-4mm}\\
}

\begin{abstract}
\noindent{\bf Purpose:}  To update and extend the Carleton Laboratory for Radiotherapy Physics (CLRP) Eye Plaque (EP) dosimetry database for low-energy photon-emitting brachytherapy sources using \ebc, an open-source EGSnrc application.  The previous database, \ec, contained datasets for the Collaborative Ocular Melanoma Study (COMS) plaques (10-22 mm diameter) with  \pa or \io seeds (\bdc-computed, 2008).  The new database, \dbc, consists of newly-calculated 3D dose distributions for 17 plaques [8 COMS, 5 Eckert \& Ziegler BEBIG, and 4 others representative of models used worldwide]  for \pac,  \ioc, and \cs seeds.  \\
\noindent{\bf Acquisition and Validation Methods:}  
Plaque models are developed with \ebc, based on published/manufacturer dimensions and material data.  The BEBIG plaques (modelled for the first time) are identical in dimensions to COMS plaques but differ in elemental composition and/or density.  Previously-benchmarked seed models are used.  Eye plaques and seeds are simulated at the centre of full-scatter water phantoms, scoring in (0.05 cm)$^3$ voxels spanning the eye for scenarios: (i) ‘HOMO’: simulated TG43 conditions;  {(ii)} ‘HETERO’: eye plaques and seeds fully modelled; {(iii)} ‘HETsi’ (BEBIG only): one seed is active at a time with other seed geometries present but not emitting photons (inactive); summation over all $i$ seeds in a plaque then yields `HETsum' (includes interseed effects).  For validation, doses are compared to those from \e and published data.\\ 
\noindent{\bf Data Format and Access:}
Data are available at \url{https://physics.carleton.ca/clrp/eye_plaque_v2}, \url{http://doi.org/10.22215/clrp/EPv2}.  The data consist of 3D dose distributions (text-based EGSnrc ``3ddose'' file format) and graphical presentations of the comparisons to previously published data. \\%
\noindent{\bf Potential Applications:}
The \db database provides accurate reference 3D dose distributions to advance ocular brachytherapy dose evaluations.  The fully-benchmarked eye plaque models will be freely-distributed with \ebc, supporting adoption of model-based dose evaluations as recommended by TG-129, TG-186, and TG-221.
\mbox{~}\vspace{-11mm}\\
\end{abstract}
\noindent {\bf Key words:} CLRP,  eye plaque, dose calculation, brachytherapy, \ebc

 \clearpage
% \vfill

\setcounter{page}{1}

\setlength{\baselineskip}{0.7cm}

\pagestyle{fancy}
\pagenumbering{arabic}
\section{Introduction}\label{intro}

Eye plaque (EP)  brachytherapy plays an important role in the treatment of intraocular cancers offering delivery of conformal doses to the tumour, with steep dose gradients sparing critical organs at risk  \cite{tg221all, tg129all, ABS-OOTF}.  Survival rates with plaque therapy have been reported to be comparable to enucleation but with eye preservation and the possibility of retaining some visual function \cite{COMS18, COMS28}.  Recent work suggests inferior clinical outcomes with proton therapy compared with eye plaque brachytherapy \cite{Li17}.

The small size of the eye combined with considerable dose gradients for eye plaque brachytherapy mean that dosimetry is critical \cite{tg221all, tg129all}.  Traditionally, dose calculations for plaques containing photon-emitting sources follow the water-based approach of Task Group (TG) 43 \cite{Ri17}.  
Model-based dose calculation algorithms (MBDCAs), including Monte Carlo (MC) simulations, promise more accurate dose evaluations by accounting for non-water treatment components (applicator, sources) and patient non-water tissues.   Various MC studies have demonstrated considerable errors incurred with the TG-43 approach for eye plaque brachytherapy dose evaluation \cite{tg221all}.  For example, MC studies focusing on modelling plaque backing/insert materials (in water phantom) report dose decreases relative to TG-43 of 11 to 40\% in the tumour, and as large as 90\% in organs at risk for the standardized plaques of the Collaborative Ocular Melanoma Study (COMS) containing \io or \pa seeds  \cite{Ri11,MR08,Th08,TR10}.    Considerable differences with TG-43 have been demonstrated for other photon-emitting plaque models as well as when modelling non-water patient/ocular anatomy \cite{Th10,Th08,Le14,Le14a}.   These discrepancies with TG-43 doses motivated both TG-129 and TG-221 to recommend  that dose evaluations accounting for the effects of the plaque backing and insert be carried out in parallel with traditional TG-43 calculations \cite{tg129all,tg221all}, in accord with the general recommendations of TG-186 on adoption of MBDCAs in brachytherapy \cite{tg186all}.  

The present work supports widespread adoption of MC dose evaluations for photon-emitting eye plaques.  We use  \ebc, a freely-distributed and open-source EGSnrc application \cite{Ch16,Th18}, to develop 17 eye plaque models: 10-24 mm diameter COMS, 12-20 mm BEBIG (manufactured by Eckert \& Ziegler BEBIG, Berlin, Germany), and four 16 mm diameter plaques representative of various models in use worldwide \cite{Le14a}.  Employing benchmarked \eb \ioc, \pac, and \cs  seed models\cite{Sa20}, we simulate the eye plaques in a water phantom to generate 3D dose distributions scored in (0.05 cm)$^3$ voxels in the eye region.  
 
These 3D dose distributions are used to completely update and extend the Carleton Laboratory for Radiotherapy Physics (CLRP) Eye Plaque Brachytherapy database that was originally published in 2008 (version 1, ``CLRP\_EPv1'': contained \bdc-calculated dose distributions for 10-22 mm COMS plaques with \pa or \io seeds \cite{Th08}).  The present article describes the CLRP\_EPv2 database \url{https://physics.carleton.ca/clrp/eye_plaque_v2} (\url{http://doi.org/10.22215/clrp/EPv2}) that contains 3D dose distributions for diverse photon-emitting eye plaques, and the \eb plaque models that will be freely-distributed with the \eb distribution (\url{https://physics.carleton.ca/clrp/egs_brachy/}).

%%%%%%%%%%%%%%%%%%%%%%%%%%%%%%%%%%%%%%%%%%%%%%%%%%%%%%%%%%%%%%%%%%%%%%
%%%%%%%%%%%%%%%%%%%%%%%%%%%%%%%%%%%%%%%%%%%%%%%%%%%%%%%%%%%%%%%%%%%%%%
%%%%%%%%%%%%%%%%%%%%%%%%%%%%%%%%%%%%%%%%%%%%%%%%%%%%%%%%%%%%%%%%%%%%%%
\section{Acquisition and Validation Methods}\label{methods}

\subsection{Monte Carlo simulations of eye plaques}\label{meth:mc}

All MC calculations are  performed with EGSnrc application \eb \cite{Ch16}  (GitHub commit hash 8166234, 2020, available at \url{https://github.com/clrp-code/egs_brachy/tree/egs_brachy_2020}). The benchmarking of \eb is documented in previous publications  \cite{Ch16,Th18,Sa20}.
Transport parameters are generally EGSnrc defaults  \cite{Ma20a} , using the low-energy default specifications distributed with \ebc.  Electron transport is not modelled. The photon energy transport cutoff is set to 1 keV. Photoelectric absorption, Rayleigh scattering, Compton scattering, and fluorescent emission of characteristic x rays are simulated. Photon cross sections are from the XCOM  database \cite{BH87}.  Dose is approximated as collision kerma, scored with a tracklength estimator in voxels with mass energy absorption coefficients (distributed with \ebc; previously calculated with EGSnrc application \g \cite{Ma20}).  The ``unrenormalized'' photoelectric cross sections are used, consistent with EGSnrc default \cite{Ma20a} (note that there is ambiguity in whether renormalized or unrenormalized Scofield photoelectric cross sections are in better agreement with experimental data \cite{ICRU90a}). 

\begin{figure}[ht]
\begin{center}
 
 \includegraphics[scale=0.5]{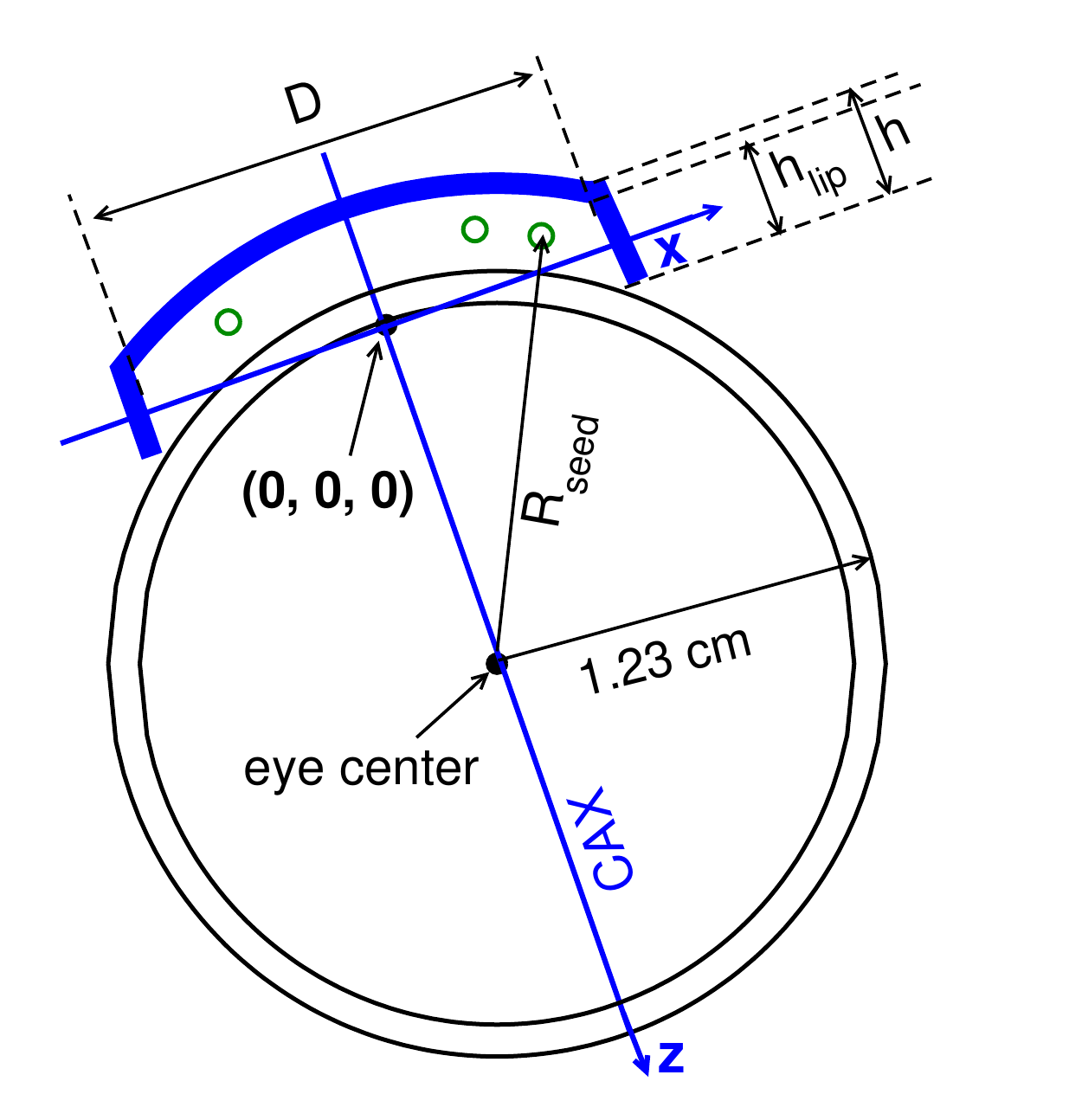}

\captionl{Schematic diagram depicting the eye plaque.  Indicated are parameters describing the eye plaque ($D$: diameter; $h$: height; $h_{lip}$: collimating lip height) and showing $R_{seed}$ as the distance from seed centre to the eye centre, as well as the coordinate system (origin at the inner sclera, 0.1~cm from idealized eye's outer surface; CAX: central axis).  

 \label{COMS16_POI_Dimen} }
\end{center}
\end{figure}

The egs$++$ class library geometry module is used to develop the eye plaque models:  COMS plaques with diameters ($D$) from 10 to 24~mm (in 2 mm increments) \cite{tg129all, Cu15a}, BEBIG plaques with diameters from 12 to 20~mm (in 2 mm increments; simulated for the first time), and four different 16 mm diameter ``representative'' plaque models  \cite{tg221all, Le14a}.  These representative plaque models were previously developed by Lesp\'erance {\it et al} \cite{Le14a} to approximate different plaque models in use worldwide (for which exact dimensions and material specifications are not widely available and/or accurately known), and include: ``Short lip-acrylic'' (Sla) \cite{Po03}, ``COMS-thin acrylic'' (Cta) \cite{Fi99b}, ``No lip-Silastic'' (NlS)\cite{Pu04}, ``Stainless steel-acrylic'' (Ssa)\cite{Gr04}.   The plaque models and associated parameters are summarized in table \ref{table:EP_sum}\mtab{table:EP_sum} with parameters/dimensions defined in figure \ref{COMS16_POI_Dimen}\mfig{COMS16_POI_Dimen}; the online \db database contains diagrams of each plaque model.  
\vspace{4mm}

\FloatBarrier
\begin{table} [h]
\begin{tiny}
\begin{center}
\caption{\sf Summary of plaques modelled: reference notation; characteristics of the backing and insert; values for diameter ($D$; does not include lip width if present), collimating lip height ($h_{lip}$, if lip present), plaque height ($h$), and radial distance from the centre of the eye to seed centres ($R_{seed}$) -- see Fig.~\ref{COMS16_POI_Dimen}.
\label{table:EP_sum}}
\begin{tabular}{lcccccccc}
\hline\hline	&		&		&		&	\multicolumn{4}{c}{Dimensions (mm)}							&		\\\cline{5-8}
Plaque model	&	Backing (thickness)	&	Insert	&	Radionuclides	&	$D$	&	$h$	&	$h_{lip}$	&	$R_{seed}$	&	Ref.	\\ \hline
COMS	&	Modulay (0.5 mm)	&	full, Silastic	&	\ioc,\pac,\cs	&	$10-24$	&	2.75	&	2.7	&	13.7	&	 \cite{tg129all, Th08, Cu15a}	\\
BEBIG	&	BioPontoStar (0.5 mm)	&	full, Silastic	&	\ioc	&	$12-20$	&	2.75	&	2.7	&	13.7	&	$^\ast$	\\
COMS - thin acrylic (Cta)	&	Modulay (0.5 mm)	&	thin 0.85 mm, acrylic	&	\ioc,\pac,\cs	&	16	&	---	&	2.7	&	13.7	&	\cite{Le14a,Fi99b}	\\
Short lip - acrylic (Sla)	&	Modulay (0.5 mm)	&	full, acrylic	&	\ioc,\pac,\cs	&	16	&	1.8	&	1.5	&	12.95	&	\cite{Le14a,Po03}	\\
No lip - Silastic (NlS)	&	Modulay (0.5 mm)	&	full, Silastic	&	\ioc,\pac,\cs	&	16	&	1.8	&	---	&	12.95	&	\cite{Le14a,Pu04}	\\
Stainless steel - acrylic (Ssa)	&	stainless steel (1 mm)	&	full, acrylic	&	\ioc,\pac,\cs	&	16	&	2.75	&	2.1	&	13.45	&	\cite{Le14a,Gr04}	\\\hline\hline\end{tabular}
\end{center}
\end{tiny}
\mbox{~}\vspace{-5mm}\\
$^*$ \scriptsize{Personal communication, Michael Andr\'assy (Eckert \& Ziegler BEBIG) June 18, 2019}\\
\vspace{4mm}
\end{table}

%\clearpage
\FloatBarrier

Each eye plaque model has a backing (with or without collimating lips), and contains either a full insert (that conforms to the outer sclera) or a thin layer of fixative that holds seeds in place.  The COMS plaques have a Modulay backing (elemental composition by mass: 77\% Au, 14\% Ag, 8\% Cu, and 1\% Pd; mass density, $\rho$ = 15.8 g/cm$^{3}$) with collimating lips and a full Silastic insert seed carrier (39.9\% Si, 28.9\% O, 24.9\% C, 6.3\% H, 0.005\% Pt; $\rho$ = 1.12 g/cm$^{3}$)\cite{tg129all}.  Compared to the COMS plaques,  the BEBIG plaques have identical geometry but with a different gold alloy comprising the backing,  ``Bio PontoStar''  (87\% Au, 10.6\% Pt, 1.5\% Zn, 0.2\% Rh, 0.2\% In, 0.2\% Ta, and 0.02\% Mn, $\rho$ = 18.8 g/cm$^{3}$), and the Silastic insert has a different mass density of ($\rho$ = 1.09 g/cm$^{3}$ [Personal communication, Michael Andr\'assy (Eckert \& Ziegler BEBIG), June 18, 2019]).  The representative plaque models (Cta, Sla, NlS) have Modulay backings, with the exception of the plaque model with the stainless steel backing (Ssa; 99.05\% Fe, 0.005\% Mn, 0.003\% Si, and 0.0015\%, $\rho$ = 7.9 g/cm$^{3}$); their inserts are either Silastic or acrylic (32\% O, 60.1\% C, 8.06\% H, $\rho$ =1.19 g/cm$^{3}$).  All representative plaques have collimating lips (but of varying lengths, $h_{lip}$ - see table \ref{table:EP_sum}) with the exception of ``No lip -Silastic'' (NlS).

All plaques models are based on fitting an idealized eye \cite{tg129all} assumed to be a sphere of radius 1.23~cm, \ie for the eye plaque models with a full insert (COMS, BEBIG, Sla, NlS, Ssa), the insert conforms to the eye's outer sclera (radius 1.23~cm).  The ``eye plaque'' coordinate system is used \cite{Th08,tg129all,Le14a} and has its origin at the inner sclera on the plaque's central axis, taken to be 0.1~cm from the outer sclera (figure \ref{COMS16_POI_Dimen}).  The plaque's central axis (CAX) defines the $z$-axis, with $x$ and $y$ coordinates transverse to the plaque.  Plaques contain between 5 and 33 seeds, with positions and orientations provided in the online \db database based on earlier publications: TG-129 report \cite{tg129all} (COMS 10 - 22 mm; BEBIG), Cutsinger {\it et al} \cite{Cu15a} (COMS 24 mm), and Lesp\'erance {\it et al} \cite{Le14a} (representative plaques: Cta, Sla, NlS, Ssa).  

Three different seeds are used for COMS and representative plaque simulations: the \pd Theragenics TheraSeed$\textsuperscript{\textregistered}$ model 200\cite{MW02,Sa20}, \io Amersham OncoSeed model 6711\cite{Do06,Sa20}, and \cs Isoray model CS-1 Rev2$\textsuperscript{\textregistered}$ \cite{Ri07,Sa20}.    For BEBIG plaques, the ophthalmic \io BEBIG IsoSeed$\textsuperscript{\textregistered}$ I-125 (I25.S16)  is used which is geometrically identical to the I25.S06 but has higher activity for use in ophthalmologic oncology \cite{He00,Sa20}.  The \eb models of these seeds were recently developed and benchmarked as part of updated 2020 CLRP TG-43v2 database \cite{Sa20}.  Photons are initialized within the seeds according to the NNDC spectra \cite{NNDC} for \pd and \cs seeds, and for \io from the NCRP report 58\cite{NCRP58}, all consistent with CLRP TG43v2 database \cite{Sa20}.  The phase space source, particle recycling and other variance reduction technique features of \eb (that enhance simulation efficiency) are not used\cite{Ch16}.  The mean energy of photons emitted from the seeds calculated by \eb are  20.51~keV (\pd model 200), 27.34~keV (\io model 6711), 30.29~keV (\cs model CS-1), and 28.16 keV (I25.S16) \cite{Sa20}.  

Plaques and seeds are modelled at the centre of a full-scatter water phantom ($\rho$ = 0.998~g/cm$^{3}$) which extends from -15 cm $\leq x,\, y,\, z \leq$ 15 cm.  Dose is scored in a $51\times51\times51$ array of (0.05 cm)$^{3}$ voxels spanning the eye region, extending from -1.275 cm $\leq x,\, y \leq$ 1.275~cm, and from -0.075 cm $\leq z\leq$ 2.475 cm (voxels centered along the plaque's central axis).  For voxels that overlap with the plaque, dose is scored only in the portion of the voxel not occupied by the plaque,  necessitating application of \ebc's voxel volume correction for which $10^9$ random points/cm$^{3}$ is used \cite{Sa20,Ch16}.

Different scenarios are simulated, all with seeds fully modelled:
\begin{enumerate}
\item {‘HOMO’}: Simulated TG-43 conditions with the plaque backing/insert modelled as water and no interseed effects (\eb in `superposition' run mode).
\item {‘HETERO’}: Eye plaques containing seeds are fully modelled in the water phantom.
\item {‘HETsi’} (BEBIG plaques only): One seed is modelled at a time, with other seed geometries present but the seeds inactive.  This produces one 3D dose distribution for each seed position in a plaque.  This is repeated for all seeds to enable superposition (accounting for seeds of possibly differing source strengths) to obtain {‘HETsum’} which is the sum of all HETsi (includes interseed effects). 
\end{enumerate}
Each calculation involves simulation of $10^{11}$ photon histories to ensure type A statistical uncertainties $\leq$ 0.2\% (1 sigma) at $z=2.26$~cm (along the central axis at the opposite side of the eye to the plaque). 
Doses are reported in terms of dose rate per unit seed air kerma strength (Gy~h${}^{-1}$~U${}^{-1}$ where 1~U$ = $1~cGy~cm$^{2}$~h${}^{-1}$) by dividing the calculated MC dose per history by the seed air kerma strength per history ($S_K^{hist}$). 
  The $S_K^{hist}$  values were previously calculated for the NIST  WAFAC detector geometry as part of the CLRP TG43v2 database  as $6.4261(6)\times 10^{-14}$~Gy\,cm$^2$/hist (\pd model 200), $3.7666(7)\times 10^{-14}$~Gy\,cm$^2$/hist (\io model 6711), $4.8312(5)\times 10^{-14}$~Gy\,cm$^2$/hist (\io I25.S16), and $3.7155(3)\times 10^{-14}$~Gy\,cm$^2$/hist (\cs model CS-1)(\url{https://physics.carleton.ca/clrp/egs_brachy/seed_database_v2})  \cite{Sa20}.  As an example of the calculation time for a clinical scenario:  simulation of a COMS 16 mm plaque fully loaded with 13 \io  seeds (model 6711) requires 200 s to achieve 2\% statistical uncertainty at the tumor apex ($z=0.5$~cm on the plaque central axis) on a single Intel Xeon 3.0 GHz CPU; calculations can be completed in a few seconds running on multiple cores (in parallel).
%on a single Intel(R) Xeon (R) CPU 3.0 GHz, to reach a 2\% statistical uncertainty at the tumour apex at $z=0.5$~cm on the central axis achieved 212 second, which is at the order of  previous calculations\cite{Ch16}.}

For validation purposes, the results of the 3D MC dose calculations (.3ddose) are imported into a CLRP in-house software tool (\url{https://physics.carleton.ca/clrp/3ddose_tools/3ddose-tools}), \eg to extract doses along different axes (CAX: central axis, $z$; transverse $x,y$ for $z=0.5,\, 1.0$~cm).

%\newpage

%%%%%%%%%%%%%%%%%%%%%%%%%%%%%%%%%
\subsection{\db data overview} \label{ EPv2 data range}
%%%%%%%%%%%%%%%%%%%%%%%%%%%%%%%%%%

\begin{landscape}
\begin{figure}[ht]
\begin{center}

\includegraphics[width=0.62\textwidth]{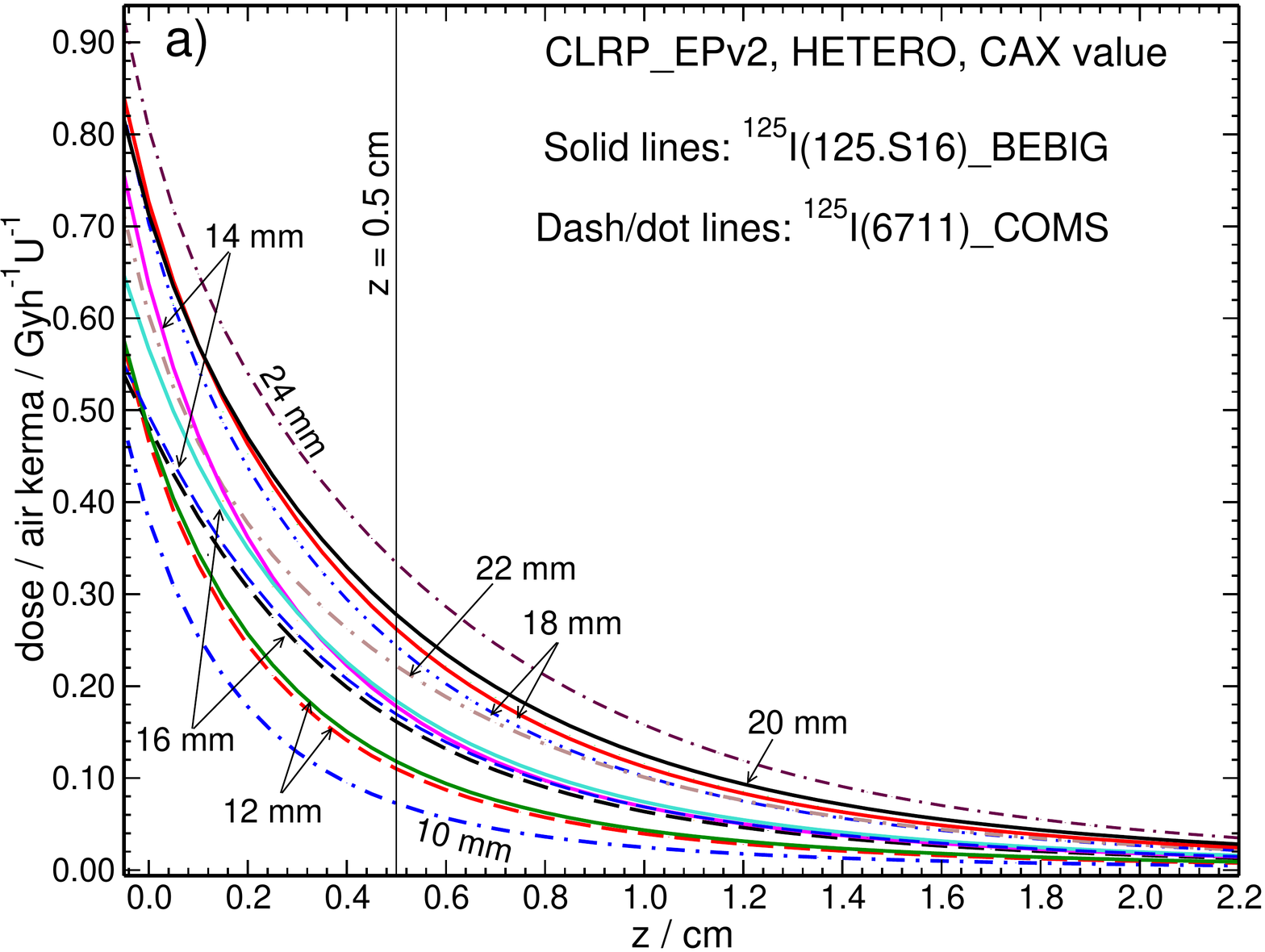}
\includegraphics[width=0.62\textwidth]{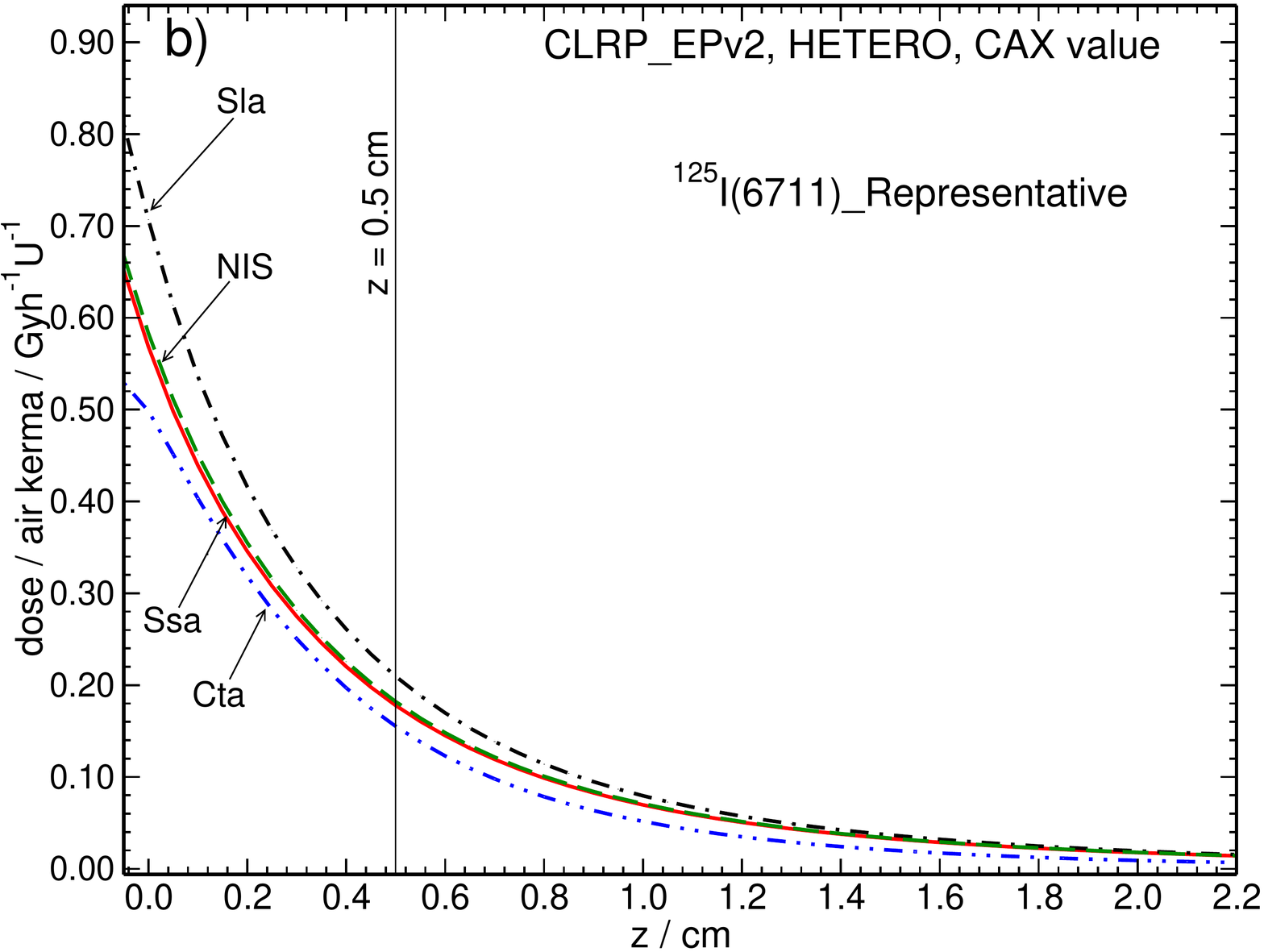}
\includegraphics[width=0.62\textwidth]{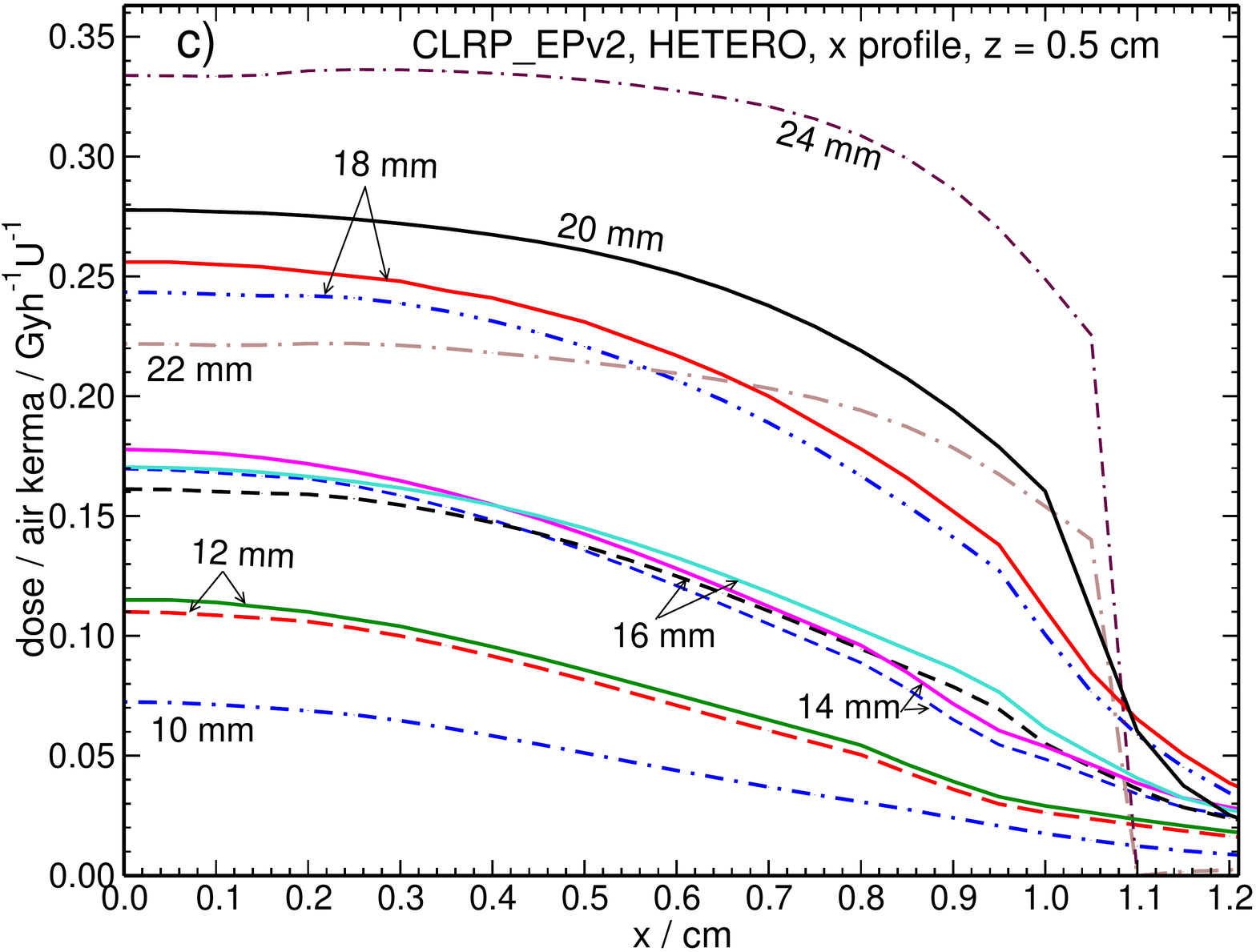}
\includegraphics[width=0.62\textwidth]{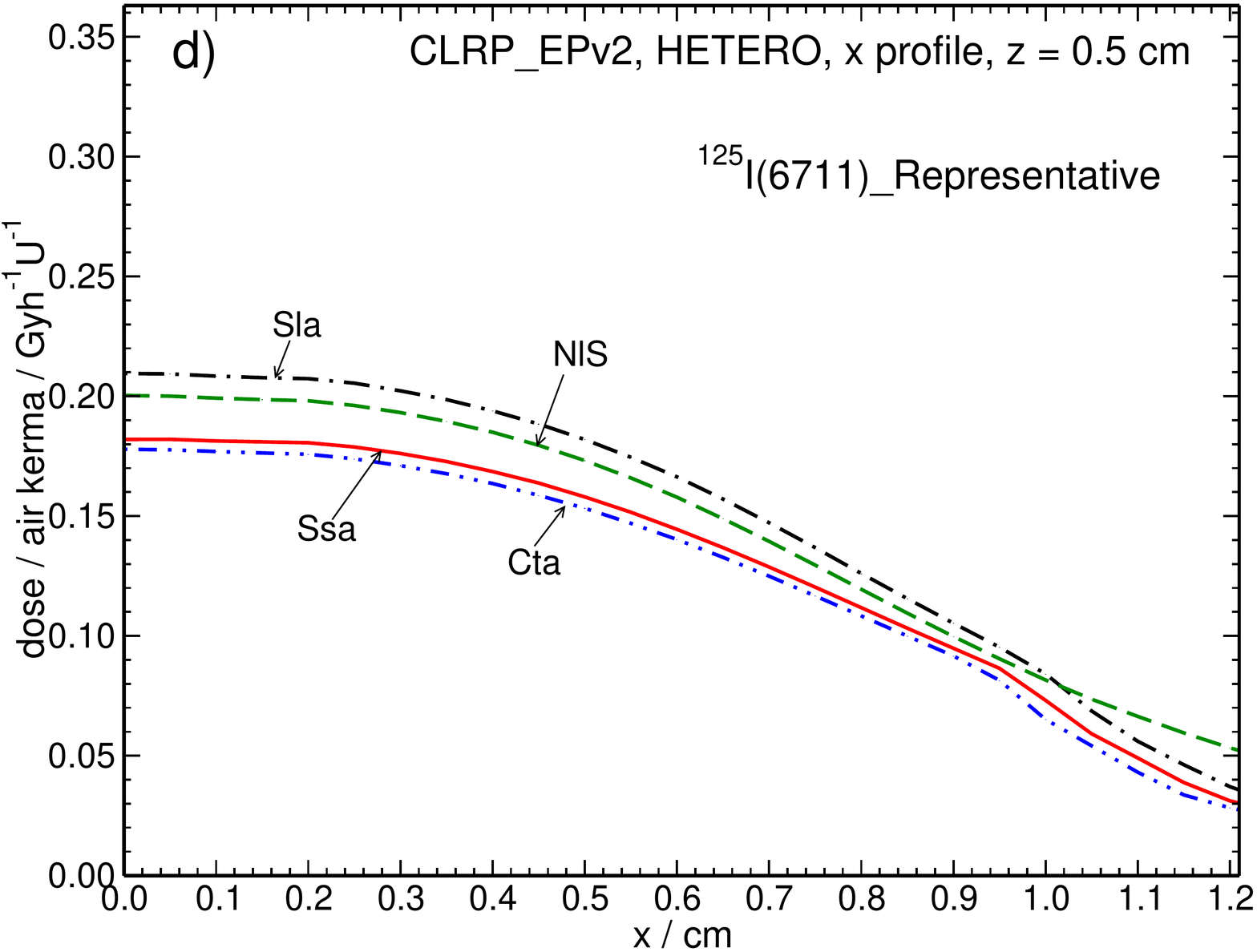}

\caption{ {Overview of \db HETERO results (for \ioc): doses along the central axis (CAX; $z$-axis) for (a) COMS and BEBIG and (b) representative plaques; doses along the transverse axis ($x$, restricted to $x\geq0$ for clarity) at $z=0.5$~cm (indicated by the vertical line in (a,b)) for (c) COMS and BEBIG and (d) representative plaques.  %COMS (dash/dot lines; model 6711), BEBIG (solid lines; I25.S16), and representative plaques (red dashed lines; 6711).
Statistical uncertainties  are $\leq$ 0.2\% along the central axis at the opposite side of the sclera.}}

\label{CLRPv2_data rangr} 
\end{center}
\end{figure}
\end{landscape}

Table~\ref{table:all_EP_v2_Cax_values_eb}\mtab{table:all_EP_v2_Cax_values_eb} and figure~\ref{CLRPv2_data rangr}\mfig{CLRPv2_data rangr} summarize \db data for all 17 plaques loaded with the different seeds, with additional data provided  online.  
In general, considerable dose  reductions for HETERO relative to HOMO doses are observed (\eg Table~\ref{table:all_EP_v2_Cax_values_eb}), due to the attenuation and scattering in the backing/insert material.  Doses decrease considerably with position along the plaque central axis, with steeper fall-off for lower energy \pd (20.51 keV) compared with higher energy \io [27.34 keV (6711), 28.16 keV (I25.S16)] or \cs (30.29 keV) seeds.  These general observations are consistent with earlier work \cite{Th08,Th10,TR10,MR08,Le14,Le14a,Ri11}.
The lowest dose per unit seed air kerma strength along central and transverse axes is for 10~mm COMS which is the smallest-diameter plaque; conversely, the highest values are for  24~mm COMS, the largest-diameter plaque (figure~\ref{CLRPv2_data rangr}).  In general, HOMO and HETERO dose rates per unit seed air kerma strength for BEBIG plaques are greater than those for the corresponding COMS plaques.

 All representative plaque models generated greater dose rates per unit seed air kerma strength compared to the 16 mm COMS or BEBIG  plaques with the lowest dose rate for Cta and highest dose rate for Sla plaque. Also, the transverse dose profiles are relatively flat (near the central axis; 0 to 0.8~cm) for plaques with higher seed number capacity (22 and 24 mm) due to the seed distribution in the plaque.\\

\begin{table} [h]
\begin{center}
\caption{\sf Summary of \db HETERO/HOMO dose ratios at depths along the central axis for each  plaque fully-loaded with \io (model 6711 or  I25.S16 for BEBIG), \pa (model 200), or \cs (CS-1 Rev2) seeds. The zero before the point is omitted (\eg $0.889 = .889$). Statistical uncertainties are less than  0.2\%.  Values of HOMO and HETERO dose rate per unit seed air kerma strength (Gy U$^{-1}$ h$^{-1}$) are provided in the \db webpage.   
\label{table:all_EP_v2_Cax_values_eb}}
\begin{tiny}% scriptsize%{footnotesize}%{small}
%\begin{tabular}{|m{3em}|ccccccccc|cccc|ccccc}
%\begin{tabular}{m{3em}cccccccccccccccccc}
\begin{tabular}{lcccccccccccccccccc}

%\hline

%\usepackage{multirow}
\hline
  \hline
%\multirow{4}{*}{Multirow} & \\
%\multirow{25}[1]{*}{Multirow} & 0.0 \bigstrut[t] \\
\multicolumn{11}{c}{ \bf{ \ ~~~~~~~~~~~~~~~~ COMS (mm) }} & \multicolumn{3}{c}{ \bf{ \ ~~ BEBIG (mm) }} & \multicolumn{5}{c}{ \bf{ \ ~~~~~~ Representative }}\\
\cmidrule(lr){3-10} 
\cmidrule(lr){11-15} 
\cmidrule(lr){16-19} 
% \cmidrule(lr){10-17}\cmidrule(lr){18-25}

 \hline

 \vspace{-0.5mm}\bf Nuc  & $z$(cm)~	&	10	&	12	&	14	&	16	&	18	&	20	&	22  &	24	&	12	&	14	&	16	&	18	 &	20  & Cta	&	Sla	&	Ssa	&	NlS		\\
\hline	
& 0.0	&	.889	&	.875	&	.867	&	.852	&	.853	&	.848	&	.854	&	.844	&	.879	&	.871	&	.857	&	.857	&	.852	&	1.00	&	1.02	&	.964	&	.922	\\

& 0.1	&	.897	&	.887	&	.880	&	.870	&	.866	&	.862	&	.863	&	.857	&	.885	&	.878	&	.869	&	.864	&	.861	&	.995	&	1.01	&	.963	&	.934	\\

&	0.2	&	.897	&	.890	&	.884	&	.878	&	.872	&	.870	&	.869	&	.864	&	.884	&	.879	&	.873	&	.867	&	.864	&	.988	&	1.00	&	.961	&	.938	\\
 \bf{\io} &	0.5	&	.883	&	.881	&	.880	&	.878	&	.874	&	.873	&	.872	&	.869	&	.871	&	.869	&	.868	&	.864	&	.862	&	.968	&	.978	&	.945	&	.935	\\
&	1.0	&	.854	&	.855	&	.857	&	.859	&	.858	&	.859	&	.859	&	.858	&	.843	&	.844	&	.846	&	.845	&	.844	&	.939	&	.949	&	.923	&	.919	\\
&	1.5	&	.832	&	.834	&	.835	&	.838	&	.838	&	.839	&	.839	&	.836	&	.824	&	.822	&	.825	&	.825	&	.824	&	.915	&	.927	&	.903	&	.904 \\

\hline

&	0.0	&	.780	&	.762	&	.753	&	.731	&	.737	&	.732	&	.741	&	.730	&	&	&	&	&	&	.989	&	1.01	&	.982	&	.850	\\
&	0.1	&	.803	&	.790	&	.781	&	.769	&	.767	&	.763	&	.766	&	.759	&	&	&	&	&	&	.984	&	.999	&	.982	&	.880	\\
&	0.2	&	.812	&	.803	&	.796	&	.788	&	.783	&	.781	&	.782	&	.776	&	&	&	&	&	&	.980	&	.992	&	.981	&	.893	\\
\bf{\pd} &	0.5	&	.814	&	.812	&	.809	&	.807	&	.804	&	.802	&	.802	&	.800	&	&	&	&	&	&	.967	&	.977	&	.973	&	.904	\\
&	1.0	&	.802	&	.803	&	.803	&	.804	&	.804	&	.803	&	.805	&	.803	&	&	&	&	&	&	.953	&	.962	&	.962	&	.901	\\
&	1.5	&	.793	&	.794	&	.793	&	.795	&	.794	&	.794	&	.795	&	.796	&	&	&	&	&	&	.943	&	.951	&	.953	&	.895	\\

\hline

&	0.0	&	.919	&	.906	&	.896	&	.885	&	.881	&	.877	&	.882	&	.871	&	&	&	&	&	&	1.00	&	1.02	&	.948	&	.937	\\
&	0.1	&	.923	&	.912	&	.904	&	.897	&	.889	&	.886	&	.888	&	.879	&	&	&	&	&	&	.992	&	1.01	&	.945	&	.944	\\
&	0.2	&	.920	&	.913	&	.905	&	.900	&	.892	&	.889	&	.890	&	.882	&	&	&	&	&	&	.984	&	.996	&	.941	&	.945	\\
\bf{\cs} &	0.5	&	.897	&	.896	&	.893	&	.893	&	.887	&	.887	&	.887	&	.881	&	&	&	&	&	&	.961	&	.970	&	.925	&	.937	\\
&	1.0	&	.859	&	.861	&	.861	&	.865	&	.863	&	.864	&	.866	&	.861	&	&	&	&	&	&	.923	&	.935	&	.896	&	.916	\\
&	1.5	&	.830	&	.832	&	.833	&	.838	&	.836	&	.838	&	.839	&	.835	&	&	&	&	&	&	.895	&	.905	&	.874	&	.897	\\

\hline
\hline

%\end{longtable}
\end{tabular}
\end{tiny}%{scriptsize}%{footnotesize}%{small}
\end{center}
\mbox{~}\vspace{-5mm}\\

\end{table}
%\end{landscape}
%\end{sidewaystable}
\typeout{***********end first table********************}

%

%%%%%%%%%%%%%%
%%%%%%%%%%%%%%%%%%%%%%

%%%%%%%%%%%%%%%%%%%%%%%%

\FloatBarrier
\FloatBarrier

%%%%%%%%%%%%%%%%%%%%%%%%%%%%%%%%%%%%

\subsection{Data validation}\label{methods:validation}

For validation, dose distributions (from 3ddose files: dose rates per unit seed air kerma strength, see section \ref{meth:mc}) are compared to previously-published doses computed with \bd data from \e for 10 - 22 mm COMS with \ioc, \pac \cite{Th08,TR10} and 16~mm COMS with \cs (Ref. \cite{Le14}); \mcnp data\cite{MR08,Ri11} for 10 - 22~mm COMS plaques (\ioc, \pac, \cscc);  \bd data for representative plaques \cite{Le14a}. It is not possible to directly compare results for BEBIG and 24~mm COMS plaques because the present study is the first published MC simulation of these plaques (but results for these plaques are compared to the other plaque sizes/types).  The following subsections summarize some of the comparisons, including doses along central and transverse axes, as well as at points of interest in the eye (on the central axis: sclera, optic disc, posterior pole, fovea, eye center, lens, lacrimal gland for 8 different plaque positions of the COMS 16 mm plaque, following the work of Rivard {\it et al} \cite{Ri11}).  Doses are compared directly and also via the the percent difference of doses for \eb ($D_{\eb}$) and another code ($D_{code}$, from \bd or \mcnpc):
 \begin{equation}
   \%\Delta(\eb,\textup{code}) = \frac{D_{\eb}-D_{code}}{\ D_\eb}    \times 100\%.
   \label{eq_Delta}
  \end{equation}
  The following subsections provide a sample of the comparisons carried out, demonstrating overall that \eb dose distributions are in excellent agreement with previously published results.

%%%%%%%%%%%%%%%%%%%%%%%

%%%%%%%%%%%%%%%%%%%%%%%
\newpage
\subsubsection{COMS plaques} \label{COMS-Representative-models}
%%%%%%%%%%%%%%%%%%%%%%%%

Dose distributions generated with \eb for COMS plaques compare well with published \bd and \mcnp data, as demonstrated by the overview of comparison results presented in this section. 

\begin{figure}[ht]
\begin{center}

 \includegraphics[width=0.496\textwidth]{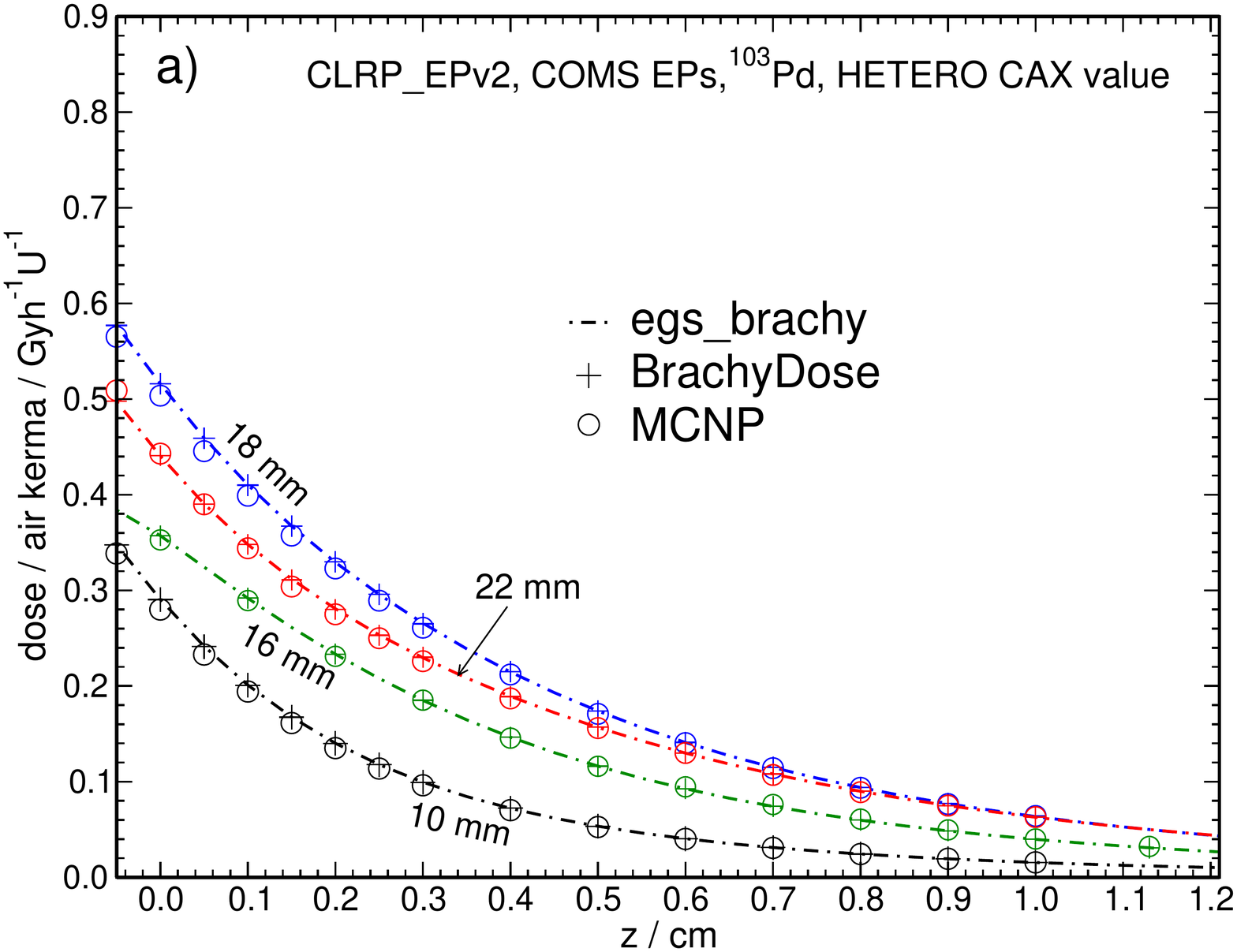}
 \includegraphics[width=0.496\textwidth]{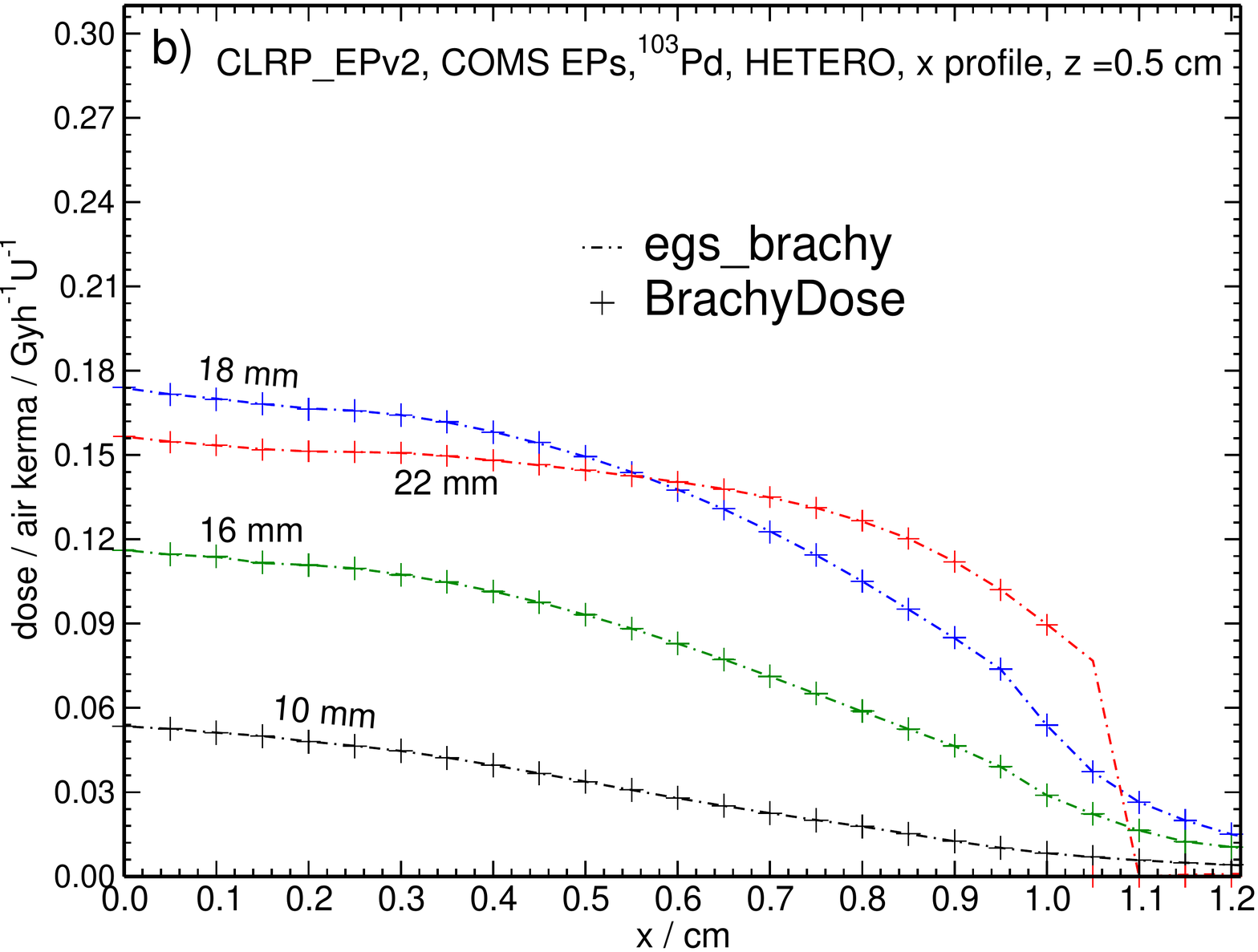}
 \includegraphics[width=0.496\textwidth]{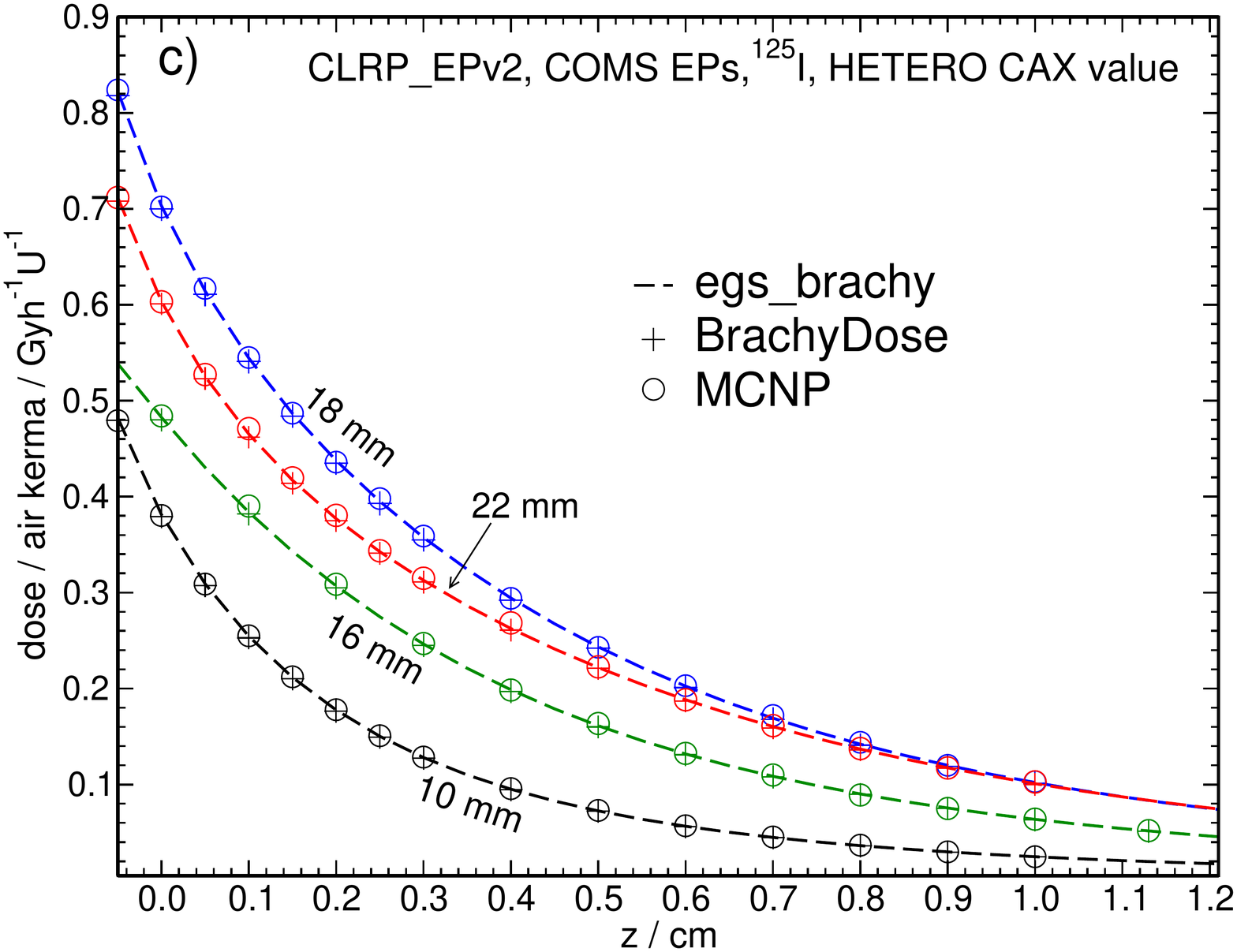}
 \includegraphics[width=0.496\textwidth]{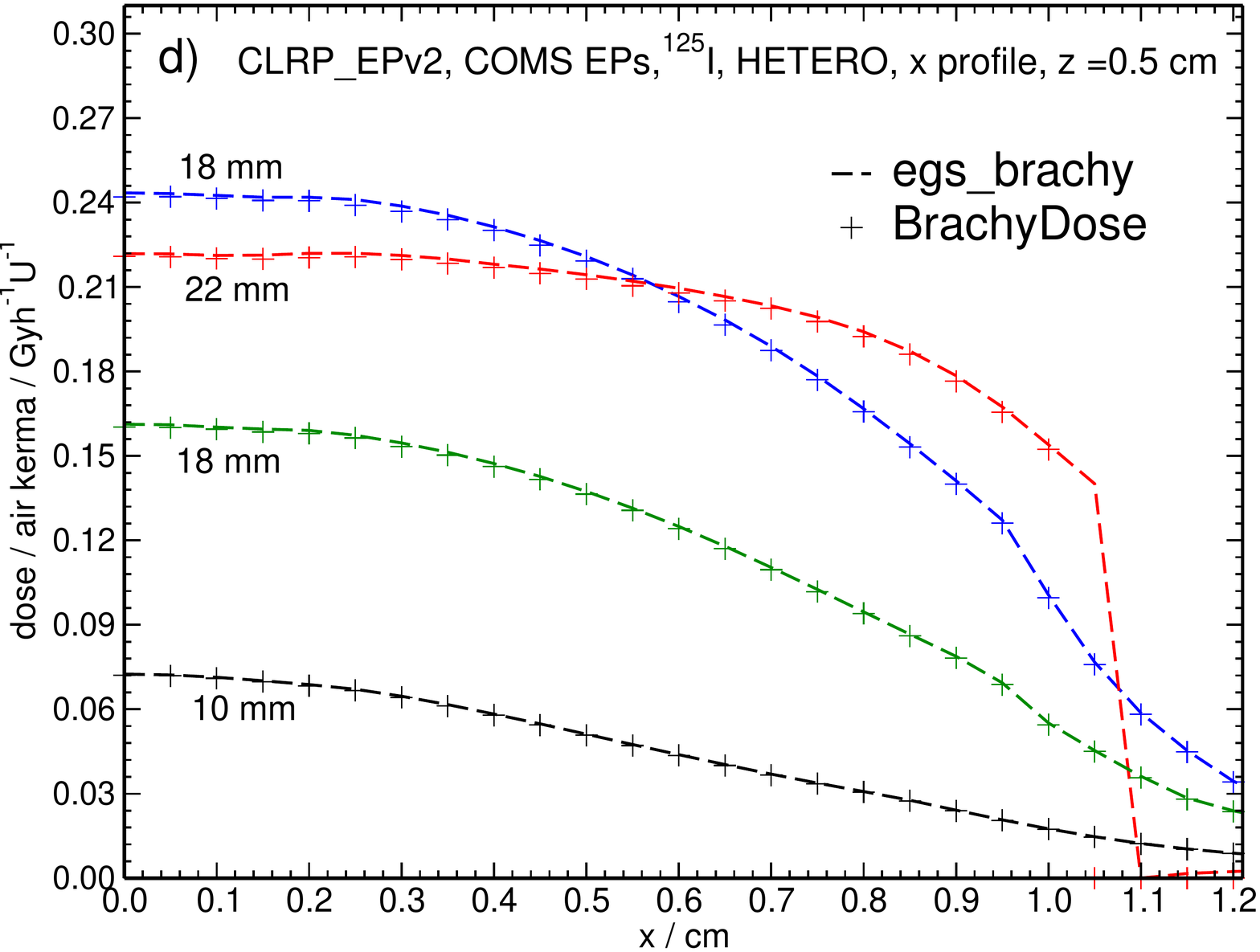}
 \includegraphics[width=0.496\textwidth]{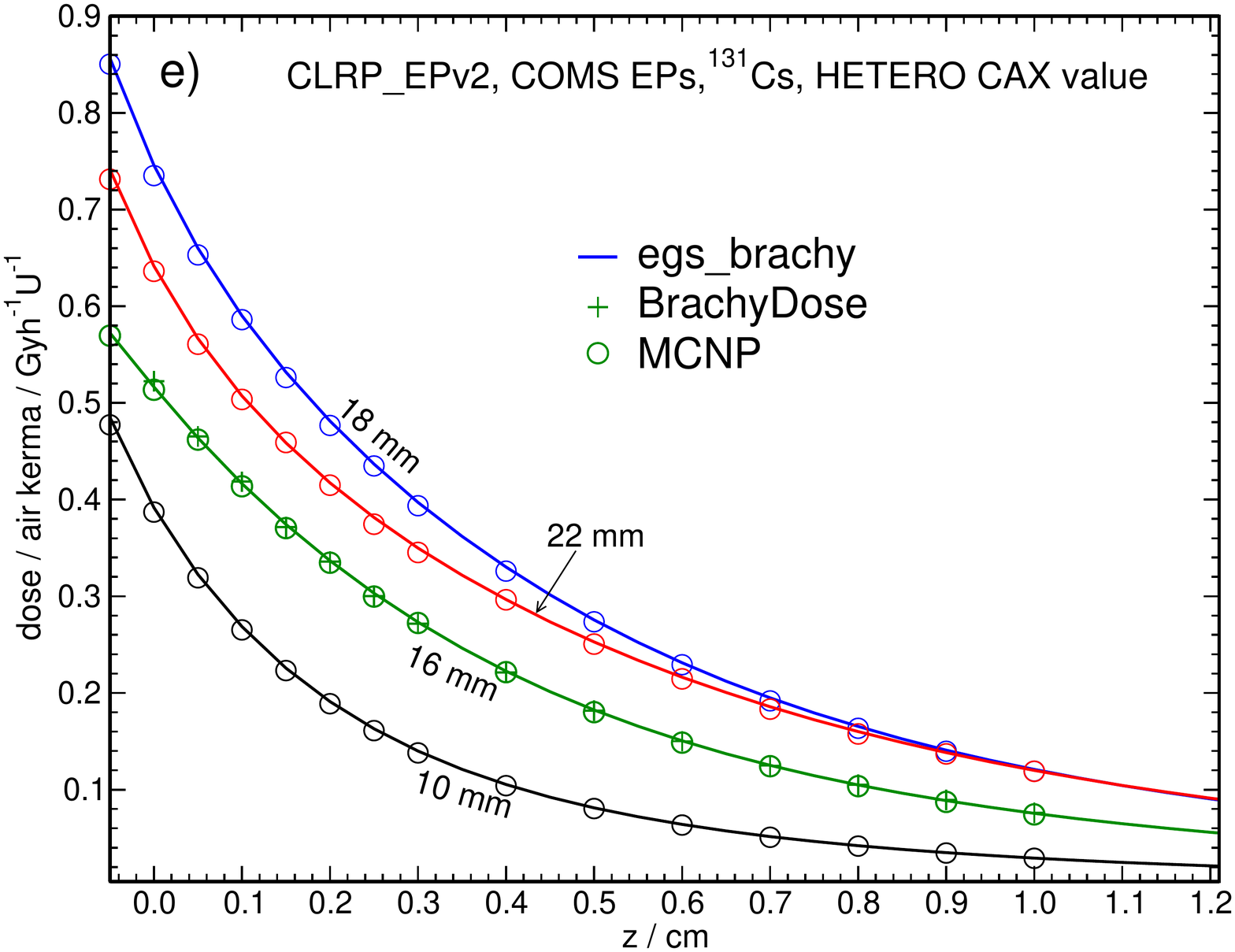}
   \includegraphics[width=0.496\textwidth]{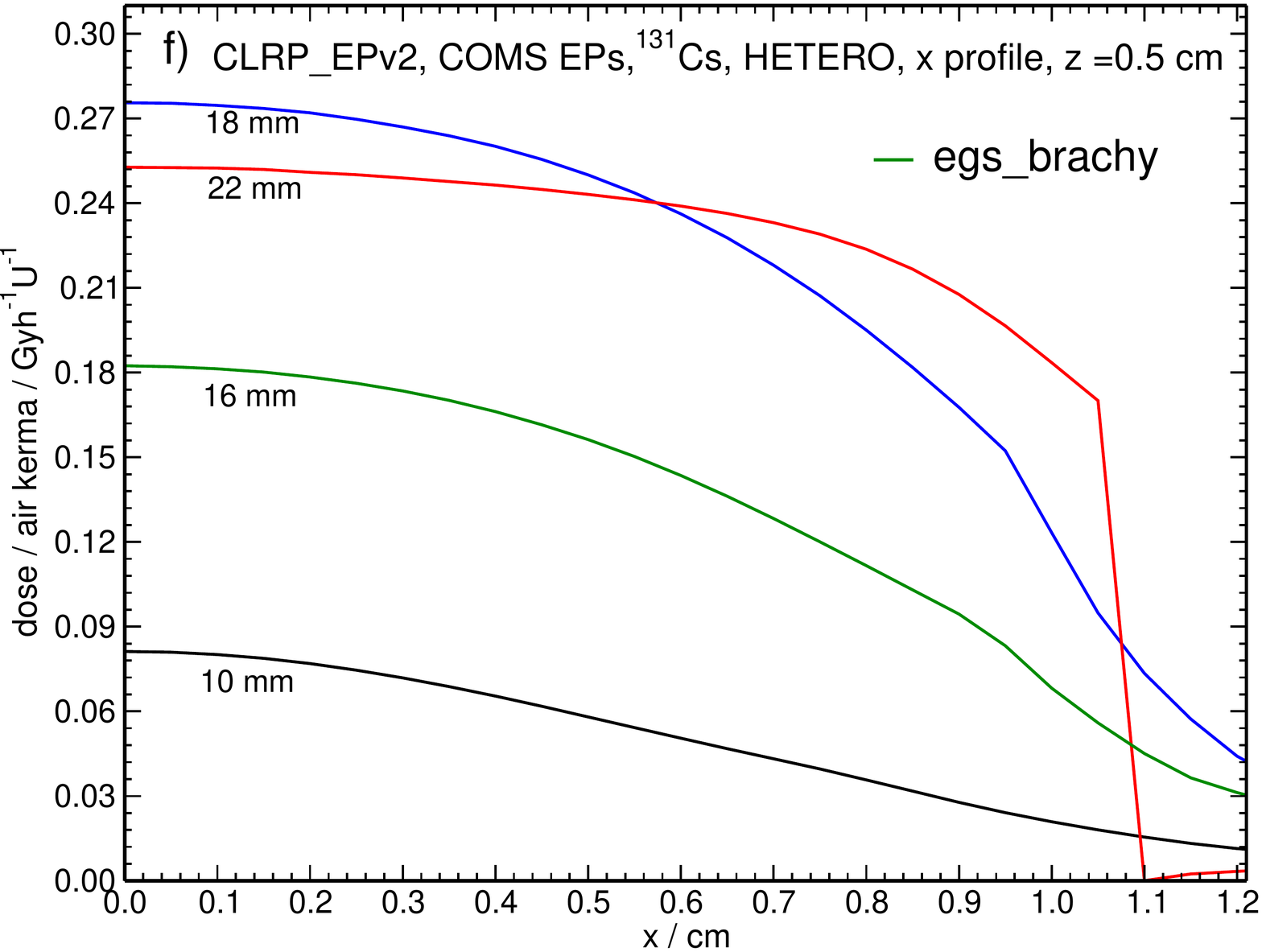}

\captionl{COMS HETERO doses: (a, c, e) along the central axis; (b, d, f) transverse $x$ axis (at 0.5 cm depth), for  \eb (lines), \BD \cite{Th08} (`$+$' symbol) and \MCNP \cite{MR08} (circle symbol) for nuclides: \pa [model 200; panel (a, b)]; \io [model 6711; panel (c,d)]; \cs [Cs-1 Rev 2; panel (e,f)]. The 10 mm, 16 mm, 18 mm, and 22 mm EPs are shown in black, green, blue, and red colors, respectively. 
 \label{CLRP_EPv2_CAX-x Dose Profile-Pd&I&cs_eb_BD} }
\end{center}
\end{figure}

\begin{figure}[ht]
\begin{center}
 
  \includegraphics[width=0.8\textwidth]{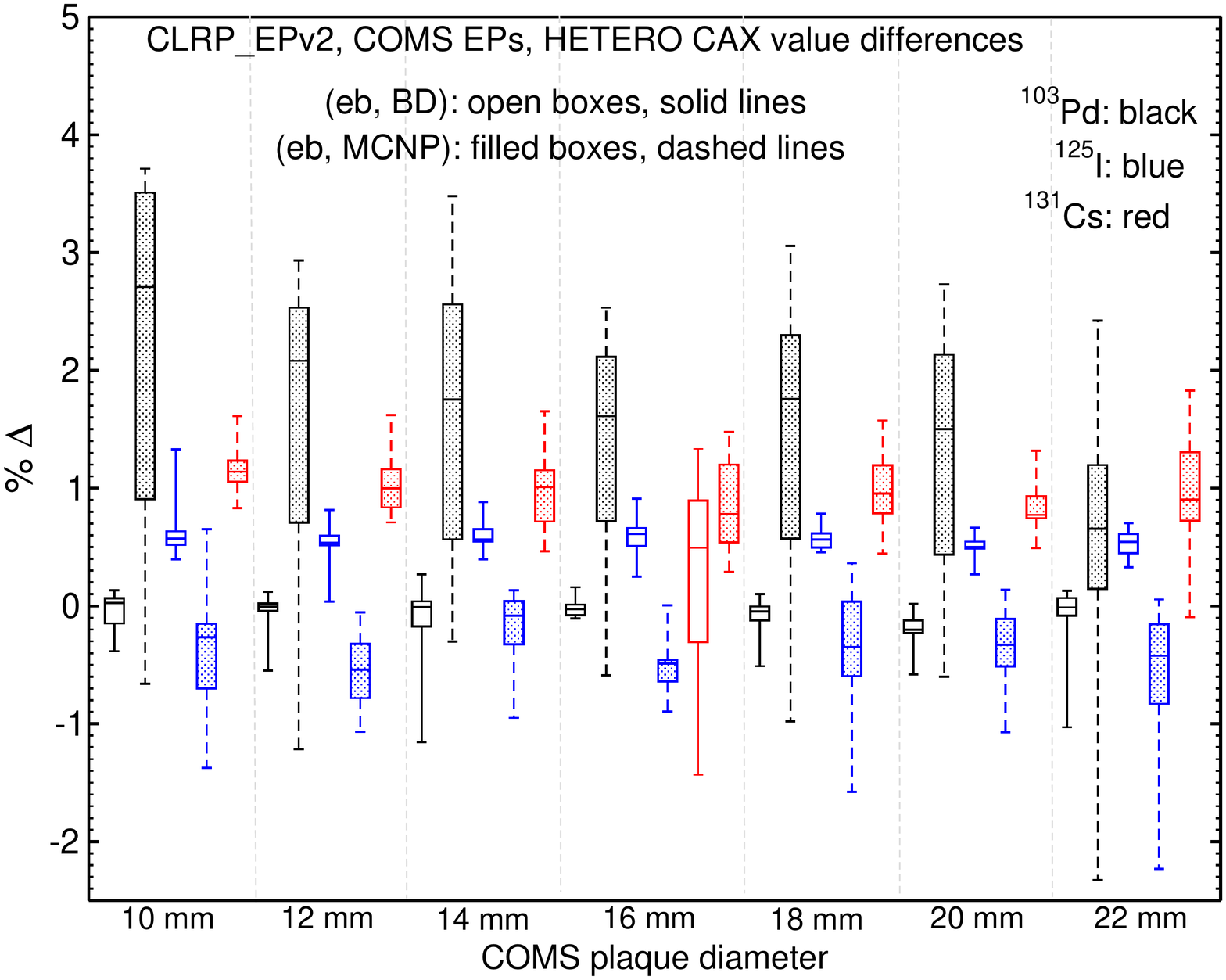}
  
\captionl{Percent  difference ($\%\Delta$; Eq.~(\ref{eq_Delta})) for HETERO doses along the central axis of COMS plaques for \eb compared with published \bd (BD) \cite{Th08,tg129all} (open boxes, solid lines) and \mcnp \cite{MR08} (filled boxes, dashed lines).  Results are shown for all plaque diameters with \pa (black), \io (blue), and \cs (red; \bd only for 16~mm because data were not published for other sizes).  Whiskers represent the total range of  $\%\Delta$ values, the boxes indicate the inner quartiles (extending between the first and third quartiles), and the horizontal line indicates the median.  Statistical uncertainties are $\leq$ 0.2\% for \ebc, $\leq$ 1\% for \bdc \cite{Th08}, and order of 2\% (7.6\% for 22~mm) for \MCNP~\cite{MR08}.  
 \label{CLRP_EPv2_box-CAX-Pd&I&cs_eb_BD-mcnp} }
\end{center}
\end{figure}

Figure~\ref{CLRP_EPv2_CAX-x Dose Profile-Pd&I&cs_eb_BD}\mfig{CLRP_EPv2_CAX-x Dose Profile-Pd&I&cs_eb_BD}  presents data for some COMS plaques along the central and transverse ($x$) axis, focusing on HETERO results; Figure~\ref{CLRP_EPv2_box-CAX-Pd&I&cs_eb_BD-mcnp}\mfig{CLRP_EPv2_box-CAX-Pd&I&cs_eb_BD-mcnp} summarizes HETERO dose differences over all COMS plaque sizes, comparing \eb with  \bd \cite{Th08,Le14a} and  \mcnp \cite{MR08}.  Across all plaque sizes,  \eb and \bd central axis HETERO doses are within 1.2\% (\pac), 1.3\% (\ioc), and 1.4\% (\cs; 16 mm COMS only), and median discrepancies are much smaller (Fig.~\ref{CLRP_EPv2_box-CAX-Pd&I&cs_eb_BD-mcnp});  agreement is comparable along transverse axes.   
Comparing \eb and \mcnp central axis HETERO doses, percent differences are at most 
3.7\% (\pac), 2.2\% (\ioc), and 1.8\% (\cscc) for HETERO. 

In considering the results in figure 4, there are different trends for $\%\Delta$ with the different radionuclides and codes.  For example, all median $\%\Delta$ values for \eb and \bd are very near zero for \pa but slightly higher and positive (near 0.5\%) for \ioc.   The values of $\%\Delta$ are generally largest for \mcnp and \ebc, with the largest and positive values for \pac, followed by \cs, whereas the values are more generally negative for \ioc.  The fact that the $\%\Delta$ comparisons lie above and below zero for different radionuclides suggests good agreement in the plaque models.  

Overall, doses along the central axis are in good agreement between \eb and previously-published \bd and \mcnp results, noting that $1\sigma$ statistical uncertainties are $<1\%$ on for \bd \cite{Th08, Le14a,Le14} and  order of 2\% for \mcnp (exception: 7.6\% for 22~mm plaque)\cite{MR08}. Discrepancies between results for the different codes may be due to a combination of  factors including cross sections, transport parameters, code versions, source spectra, mass-energy absorption coefficient values.  Melhus and Rivard \cite{MR08} estimated that the total uncertainty on their results is 5\% at a depth of 5~mm from the plaque (all three radionuclides and COMS plaque sizes, including non-statistical sources of uncertainty). 

\FloatBarrier
\FloatBarrier

%%%%%%%%%%%%%%%%%%%%%%%%%%%%%Box-whisker-CLRP-EPv2-CAX%%%%%%
\FloatBarrier
\FloatBarrier

\begin{figure}[ht]
\begin{center}
 \includegraphics[width=0.7\textwidth]{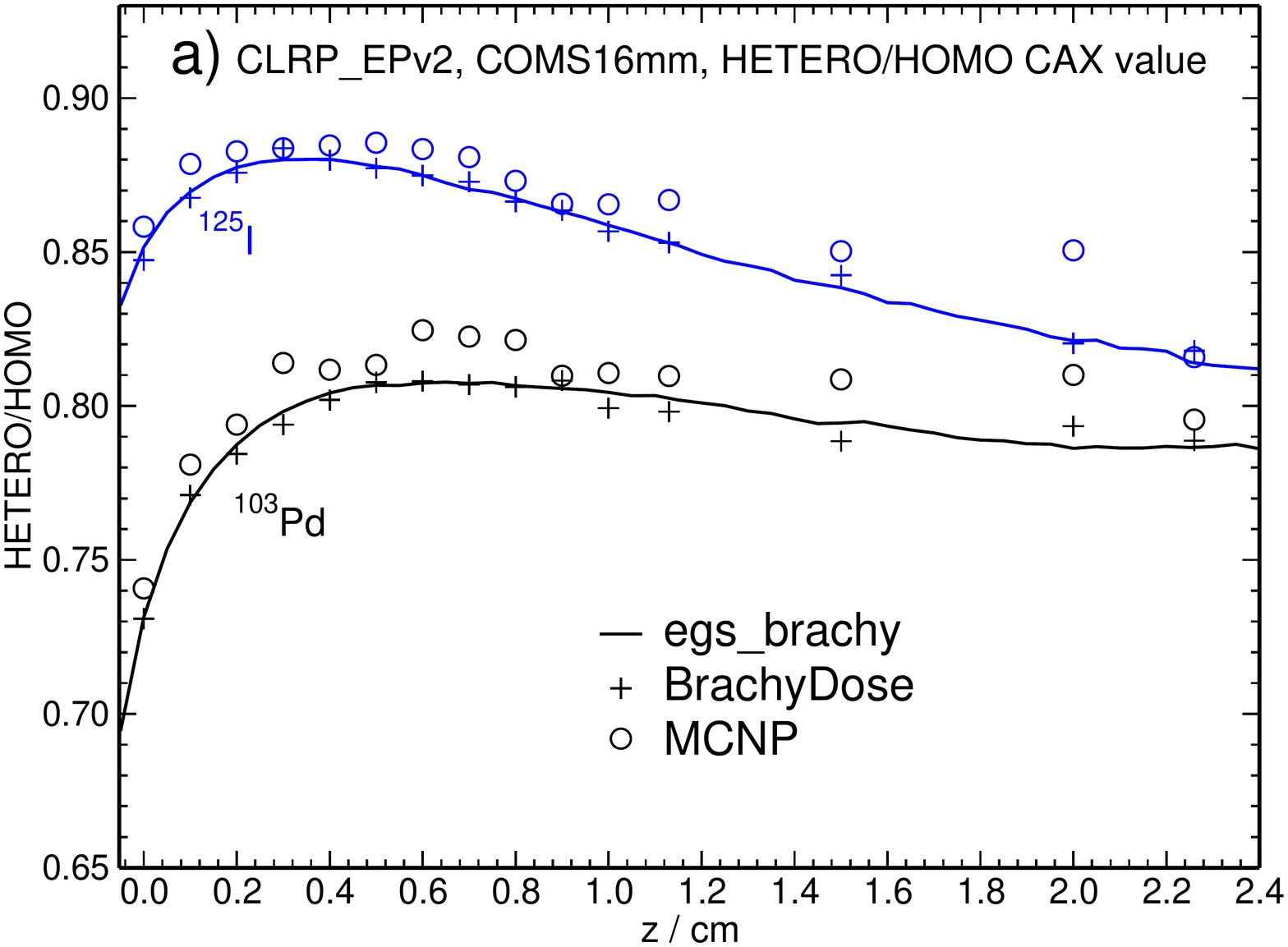}
  \includegraphics[width=0.7\textwidth]{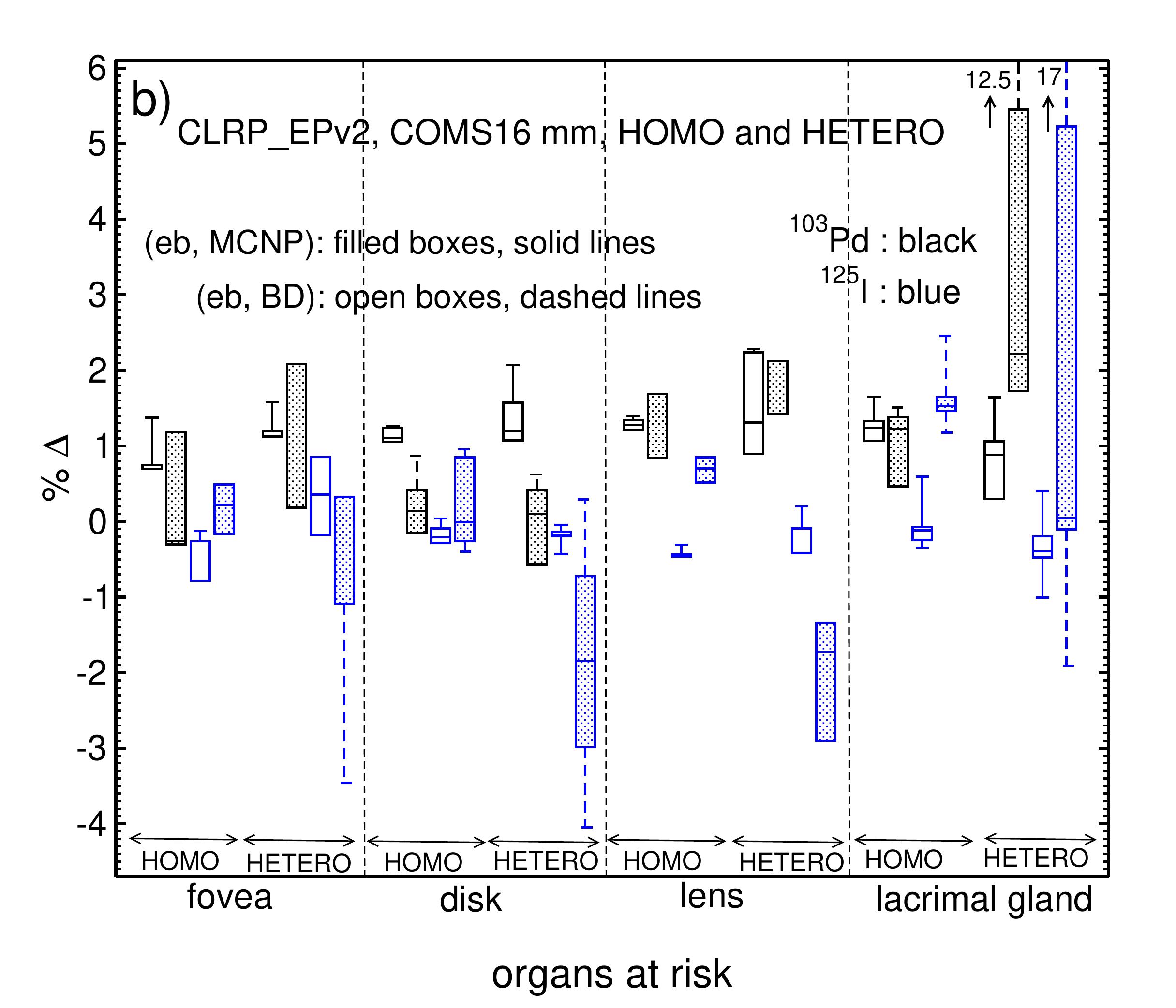}
\captionl{Comparisons of results for 16~mm COMS plaques for (a) HETERO/HOMO ratio of doses along the central axis (for \pa (200), \io (6711), \cs (CS-1)) and (b) box-and-whisker summary of percent dose differences (\eb (eb) compared with \bd (BD) and \mcnpc) over eight different plaque positions considering four organs at risk and \pd (200) and \io (6711) seeds \cite{Ri11,Th08}.

 \label{COMS16_box-CAX+POI-Pd&I} }
\end{center}
\end{figure}

Figure~\ref{COMS16_box-CAX+POI-Pd&I}\mfig{COMS16_box-CAX+POI-Pd&I} provides further comparison of \eb results with \bd and \mcnp for 16 mm COMS.  Fig.~\ref{COMS16_box-CAX+POI-Pd&I}a presents the ratio of doses HETERO/HOMO along the central axis for \ebc, \bd \cite{Th08,Le14}, and \mcnp \cite{Ri11}: comparing with \ebc, \bd ratios are within 1.5\% (\pac), 0.5\% (\ioc), and those for \mcnp are within 3.0\% (\pac), and  3.5\% (\ioc).  
The fact that dose ratios are systematically higher for \mcnp relative to \eb (and \bdc) may be partially attributed to the fact that HOMO \mcnp simulations include interseed attenuation (those for \eb and \bd do not), thus lower the dose in the denominator and increasing the HETERO/HOMO ratio.  Previous research assessed the magnitude of interseed effects, reporting that doses differed by less than 0.5\% (HETERO) and by 1 to 2\% for seeds in water (HOMO) \cite{Th08}. 

For all codes, the largest discrepancies are observed at the opposite side of the eye to the plaque where the statistical uncertainties are largest.  Figure~\ref{COMS16_box-CAX+POI-Pd&I}b summarizes comparisons of doses at organs at risk (positioned off the central axis), considering eight different plaque positions as done by Rivard {\it et al} \cite{Ri11}: differences are $\le$ 2.3\% for \eb and \bd and $\le$ 4\% for \eb and \MCNP~  with the exception of the lacrimal gland HETERO dose for two (of the eight) plaque positions considered.  For those positions, the lacrimal gland HETERO doses are very low (3.5~Gy for \pa and 6.6~Gy for \ioc, compared with 85~Gy as the prescription dose) and  \mcnp differs by 12.5\% (\pac) and 17\% (\ioc) from \ebc.  This large discrepancy in the dose at the lacrimal gland for these plaque positions has been observed previously in comparing \bd and \mcnp doses \cite{Ri11} and elsewhere in comparing \ebc, \bdc, and \mcnp results  \cite{Ri11,Th18}.

The results shown in this subsection demonstrate that there is generally closer agreement between the two EGSnrc-based code, \eb and \bdc, than between \eb and \mcnpc, consistent with previous work \cite{Th18}.  Overall, these comparisons validate the new \eb models and 3D dose distributions for COMS plaques.

%%%%%%%%%%%%%%%%%%%%%COMS16mm-HEREO/HOMO and Box-Whisker%%%%%%%%%%%%

\FloatBarrier
\FloatBarrier
%%%%%%%%%%%%%%%%%%%%%%%%%%%%%%%%%%%%%%%%%%%%%

%%%%%%%%%%%%%%%%%%%%%%%%%%%%%%

%%%%%%%%%%%%%%%%%%%%%%%%%%%%%

\subsubsection{BEBIG plaques}\label{BEBIG_EPs}

The present work offers the first MC simulations of the BEBIG model plaques, and so comparisons to previously published results are not possible.  However, the BEBIG plaques are of the same (geometric) dimensions to the COMS plaques, and thus the validation of our \eb models for those plaques (section \ref{COMS-Representative-models}) supports the new \eb BEBIG models.  The dose distributions are expected to be different for BEBIG compared with COMS due to differences in plaque backing medium (elemental composition and density), insert density, and seed model (\ioc: IsoSeed$\textsuperscript{\textregistered}$ I25.S16  for BEBIG, model 6711 for COMS) on the basis of earlier work investigating backing composition \cite{Th08} and seed type \cite{TR10}; this is confirmed by the results (\eg Table \ref{table:all_EP_v2_Cax_values_eb}).  \vspace{4mm}

\begin{figure}[ht]
\begin{center}

 \includegraphics[width=0.7\textwidth]{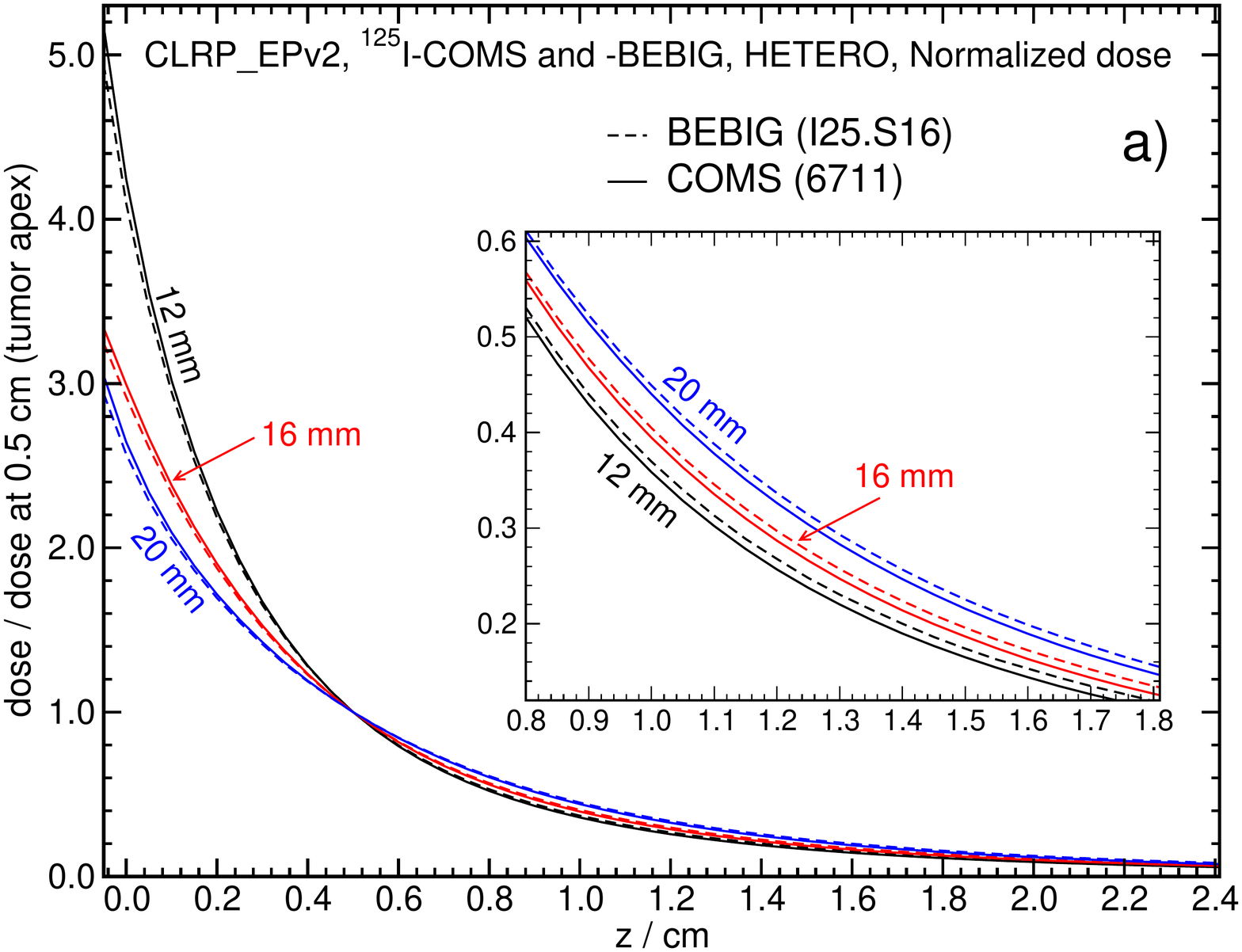}
 \includegraphics[width=0.7\textwidth]{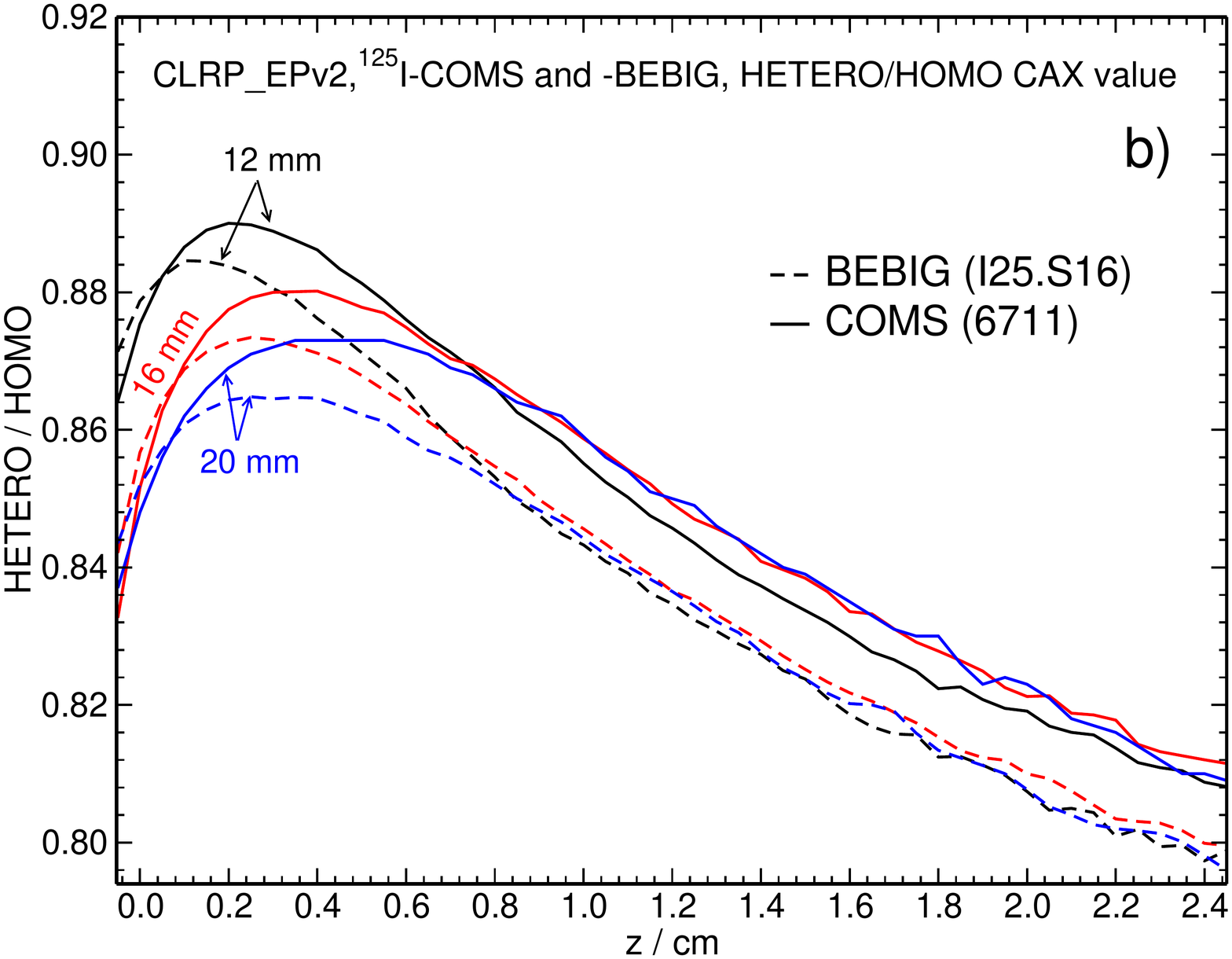}
\captionl{Summary of results for BEBIG plaques with \io model I25.S16 seeds (dashed lines) compared with corresponding COMS plaques (containing \io model 6711 seeds) for 12, 16, and 20~mm plaque diameters for doses along the central axis (a) normalized to unity at $z=0.5$~cm and (b) ratio HETERO/HOMO.  Combined statistical uncertainties for HETERO/HOMO dose ratios are $\le$ 0.2\%. 

 \label{COMS-BEBIG-HETERO/HOMO-12-16-20mm} }
\end{center}
\end{figure}

Figure~\ref{COMS-BEBIG-HETERO/HOMO-12-16-20mm}\mfig{COMS-BEBIG-HETERO/HOMO-12-16-20mm} provides a subset of dose results for a few BEBIG plaque diameters, contrasted with corresponding COMS plaques.  Fig.~\ref{COMS-BEBIG-HETERO/HOMO-12-16-20mm}a demonstrates that when the doses are normalized to the dose at $z=0.5$~cm on the plaque central axis (as an example prescription point), doses on the plaque central axis within the tumour ($z<0.5$~cm) are slightly lower for BEBIG than for the corresponding COMS plaque (by less than 2\% on average), and at the inner sclera the doses for BEBIG plaques are lower by 3.5\% (12~mm), 3.4\% (14~mm), 2.5\% (16~mm), 2.9\% (18~mm), and 2.8\% (20~mm). 

\FloatBarrier
\FloatBarrier

Figure~\ref{COMS-BEBIG-HETERO/HOMO-12-16-20mm}b presents the ratio of HETERO/HOMO doses for a subset of BEBIG and COMS plaque sizes (\io seed models: I25.S16 for BEBIG, 6711 for COMS), characterizing the heterogeneity effect due to the plaque.  Over all plaque sizes, the dose ratio HETERO/HOMO is larger for BEBIG relative to COMS very near the plaque, but further from the plaque the dose decrease is more substantial for BEBIG than COMS.  These differences between BEBIG and COMS dose distributions result from a combination of effects due to plaque backing materials of different elemental compositions and densities, different insert mass densities, and distinct seed models.
  
The lower mass density of Silastic combined with the slightly higher average energy of the BEBIG I25.S16 seeds  results in a slightly higher HETERO/HOMO dose ratio very near the plaque.  Further away, the higher atomic number of predominant plaque components (Au, Pt) for BEBIG plaques results in larger dose decreases compared with COMS plaques (no Pt; Ag with Au, among other elements), in accord with previous results comparing pure gold backings with Modulay \cite{Th08}.  Overall, the differences between HETERO/HOMO dose ratios for BEBIG and COMS are 2\% or less, consistent with previous comparisons of plaque backings \cite{Th08} and seed models \cite{TR10}.

In a separate set of simulations, interseed attenuation is investigated for the BEBIG plaques (containing  I25.S16 seeds) by comparison of HETERO simulations with and without interseed effects (\eb `normal' and `superposition' run modes, respectively).  Interseed attenuation reduces doses by $0.5\%$ (12 mm), 0.4\% (14 mm), 0.3\% (16 mm), 1.0\% (18 mm), and 1.6\% (20 mm) at the inner sclera and by $0.3\%$ (12 mm), 0.2\% (14 mm), 0.2\% (16 mm), 0.2\% (18 mm), and 0.5\% (20 mm) at $z=1$~cm.  These relatively small differences between HETERO simulations with and without interseed effects are in accord with previous work that considered other seed models \cite{Th08,TR10}.

\begin{figure}[ht]
\begin{center}

 \includegraphics[width=0.7\textwidth]{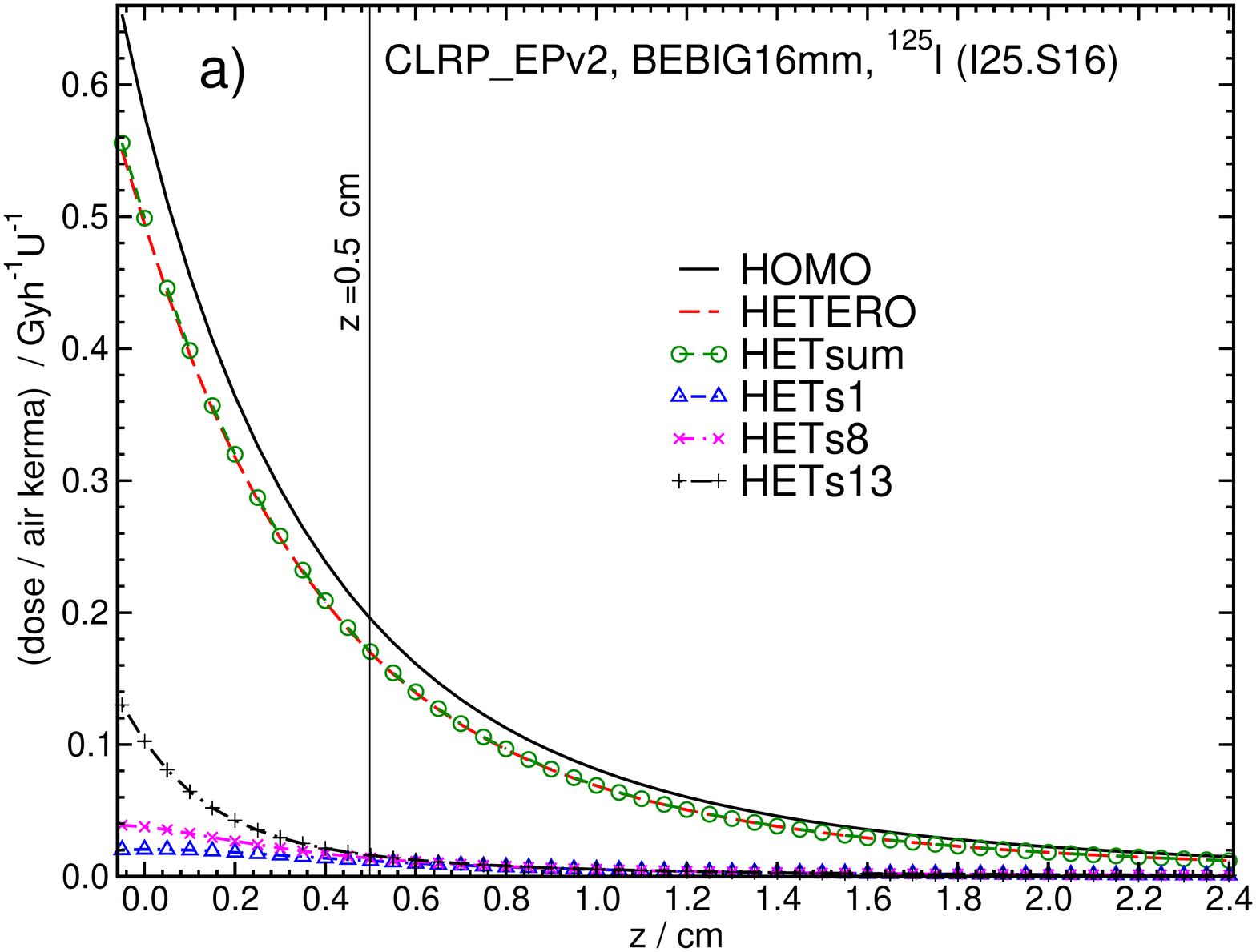}\\
 
 \includegraphics[width=0.7\textwidth]{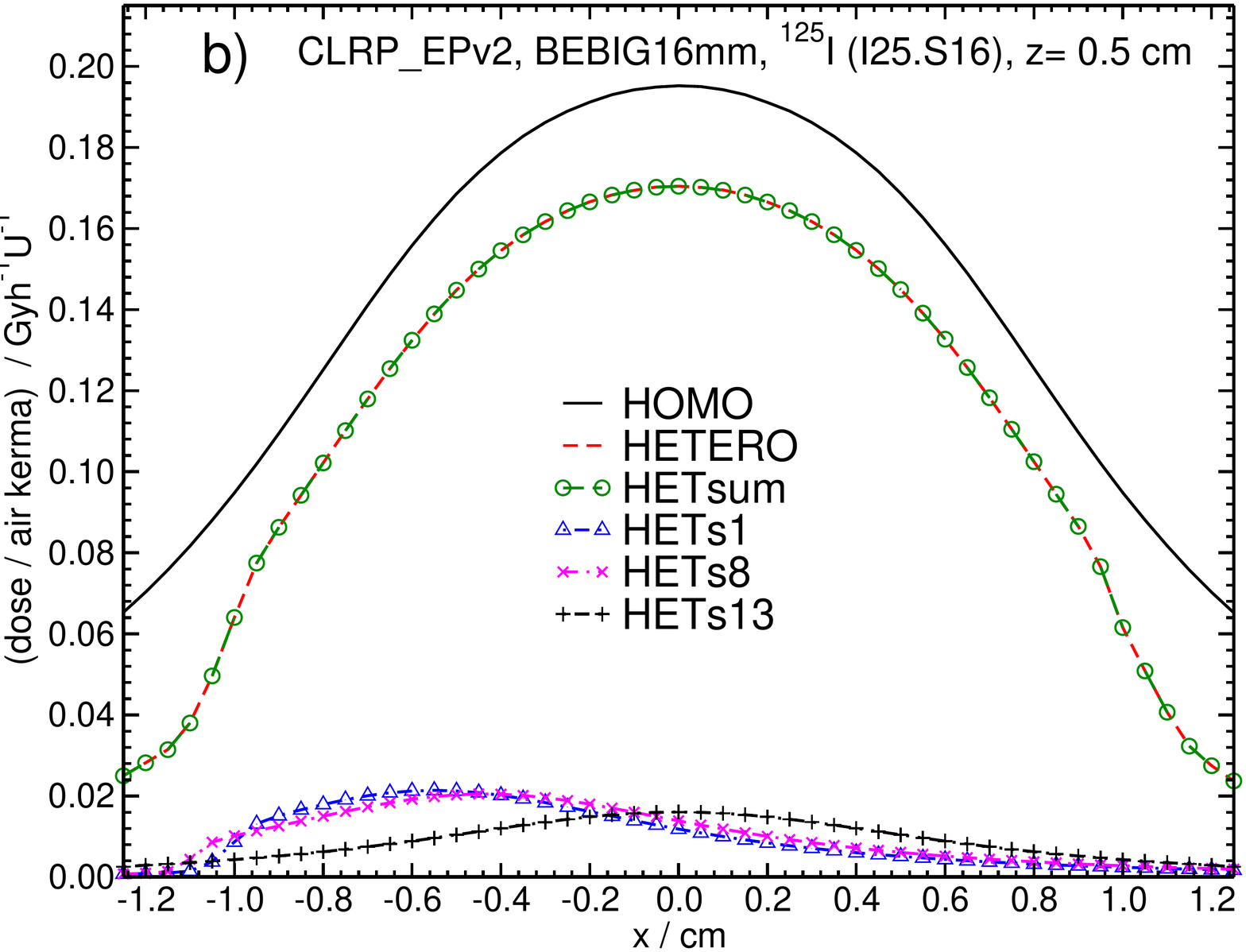}

\captionl{Summary of 16~mm BEBIG  (I25.S16) doses per unit seed air kerma strength for HOMO, HETERO, HETsum, and HETsi scenarios for (a) central axis and (b) transverse ($x$) axis at $z=0.5$~cm.  Note that central axis curves are identical for HETs1 to HETs7,  HETs8 to HETs11, and HETs12 to HETs13, because the radial distance from the seed centre to the $z$ axis is the same in each case.  Statistical uncertainties are $\leq$ 0.2\% along the central axis.

 \label{BEBIG16mm-HOMO_HET-HETsum_HETsi-cax-Xview} }
\end{center}
\end{figure}

Figure~\ref{BEBIG16mm-HOMO_HET-HETsum_HETsi-cax-Xview}\mfig{BEBIG16mm-HOMO_HET-HETsum_HETsi-cax-Xview} summarizes the HETsi and HETsum results in comparison with HETERO and HOMO.  The figures demonstrate that superposition of the HETsi results to determine HETsum provides dose distributions in excellent agreement (as expected) with HETERO results.

%%%%%%%%%%%%%%%%%%%%%%%%%%%BEBIG CAX and X profile%%%%%%%
\FloatBarrier
\FloatBarrier
%%%%%%%%%%%%%%%%%%%%%%%%%%%

%%%%%%%%%%%%%%%%%%%%%BEBIG_COMS-comapre%%%%%%

%%%%%%%%%%%%%%%%%%%%%%%%%%%%%%%%%%%%%%%%%%%%%%%%%%%%

%%%%%%%%%%%%%%%%%%%%%%%
\subsubsection{Representative plaques} \label{Representative-models}
\begin{figure}[ht]
\begin{center}

\includegraphics[width=0.54\textwidth]{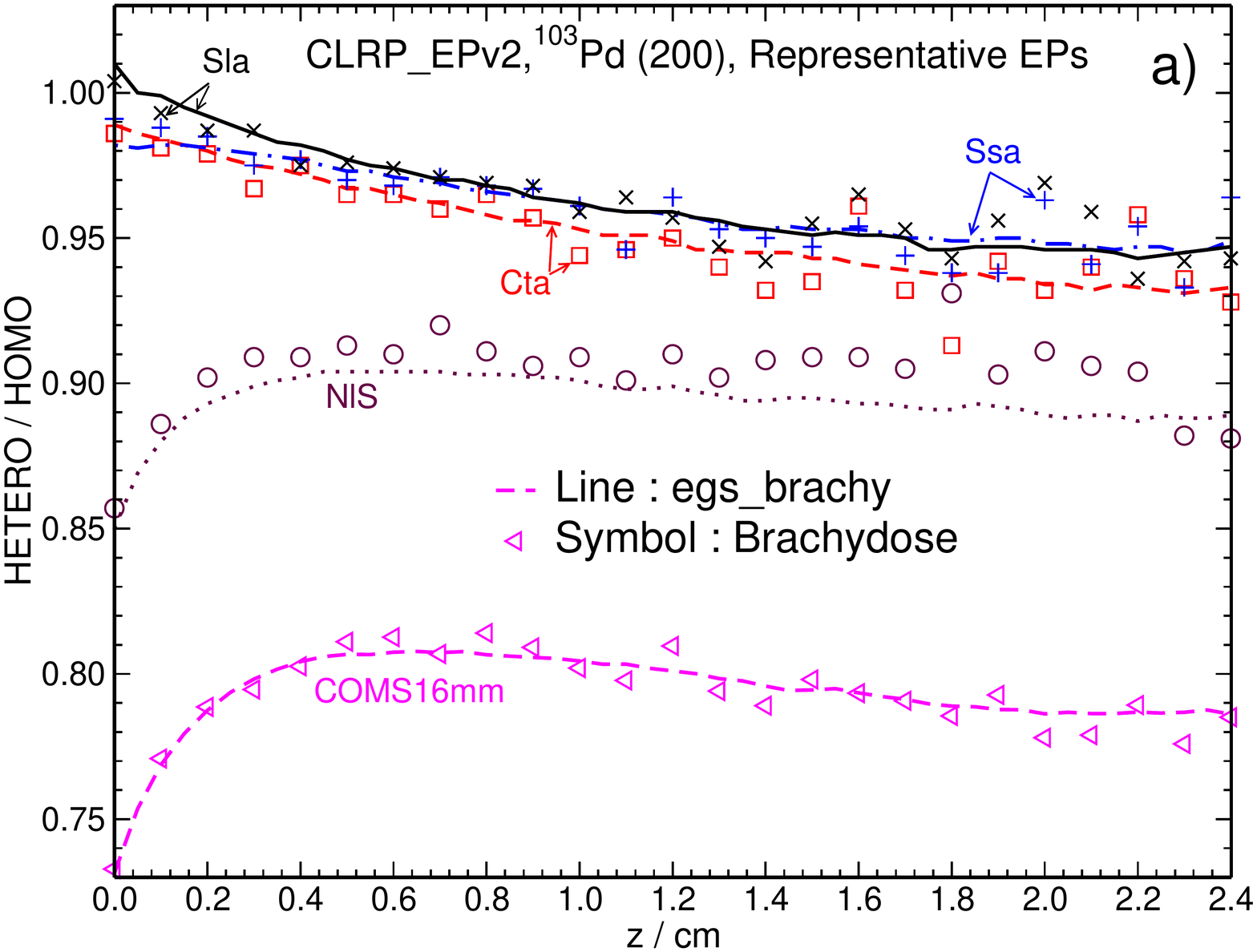}\\
\includegraphics[width=0.54\textwidth]{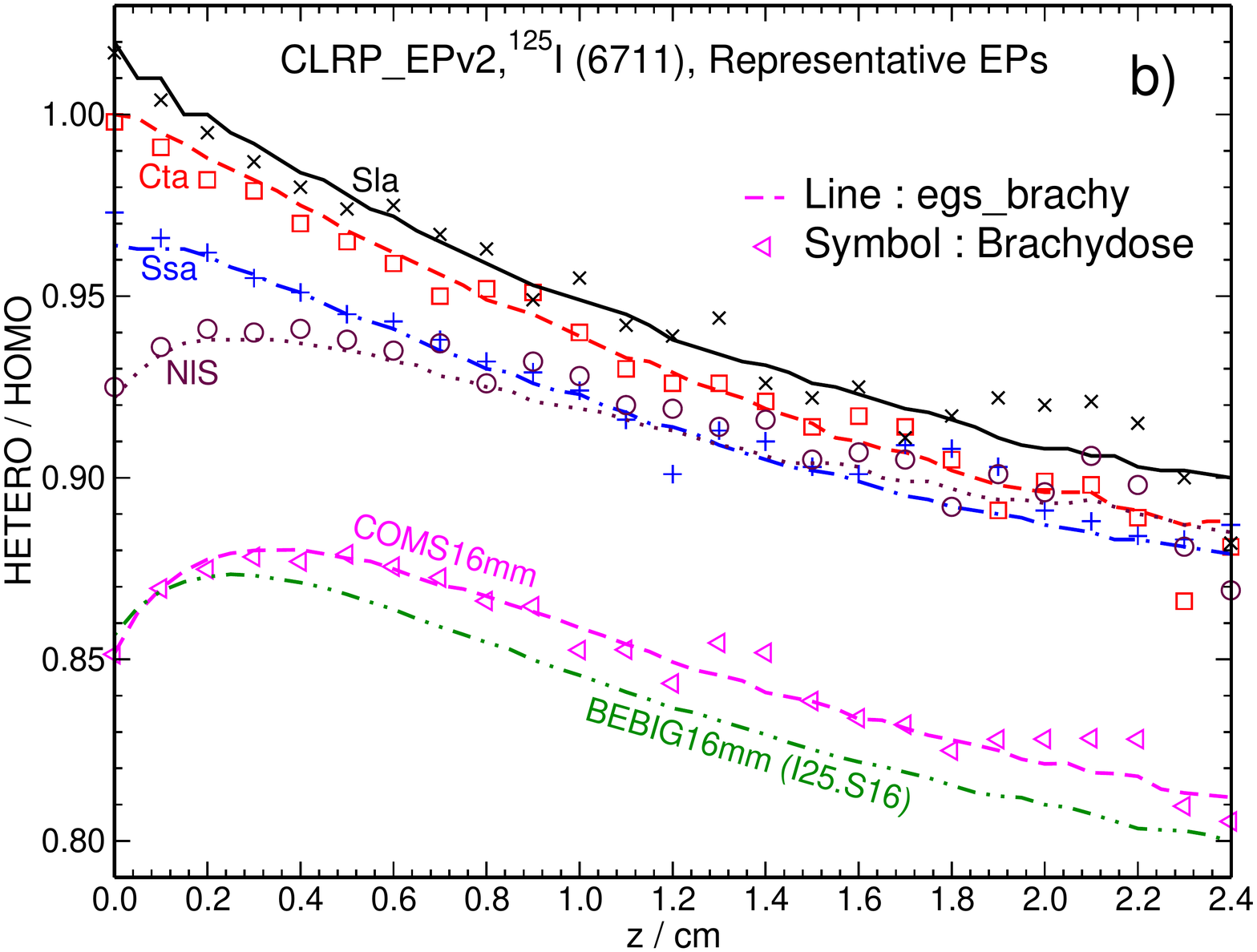}\\
\includegraphics[width=0.54\textwidth]{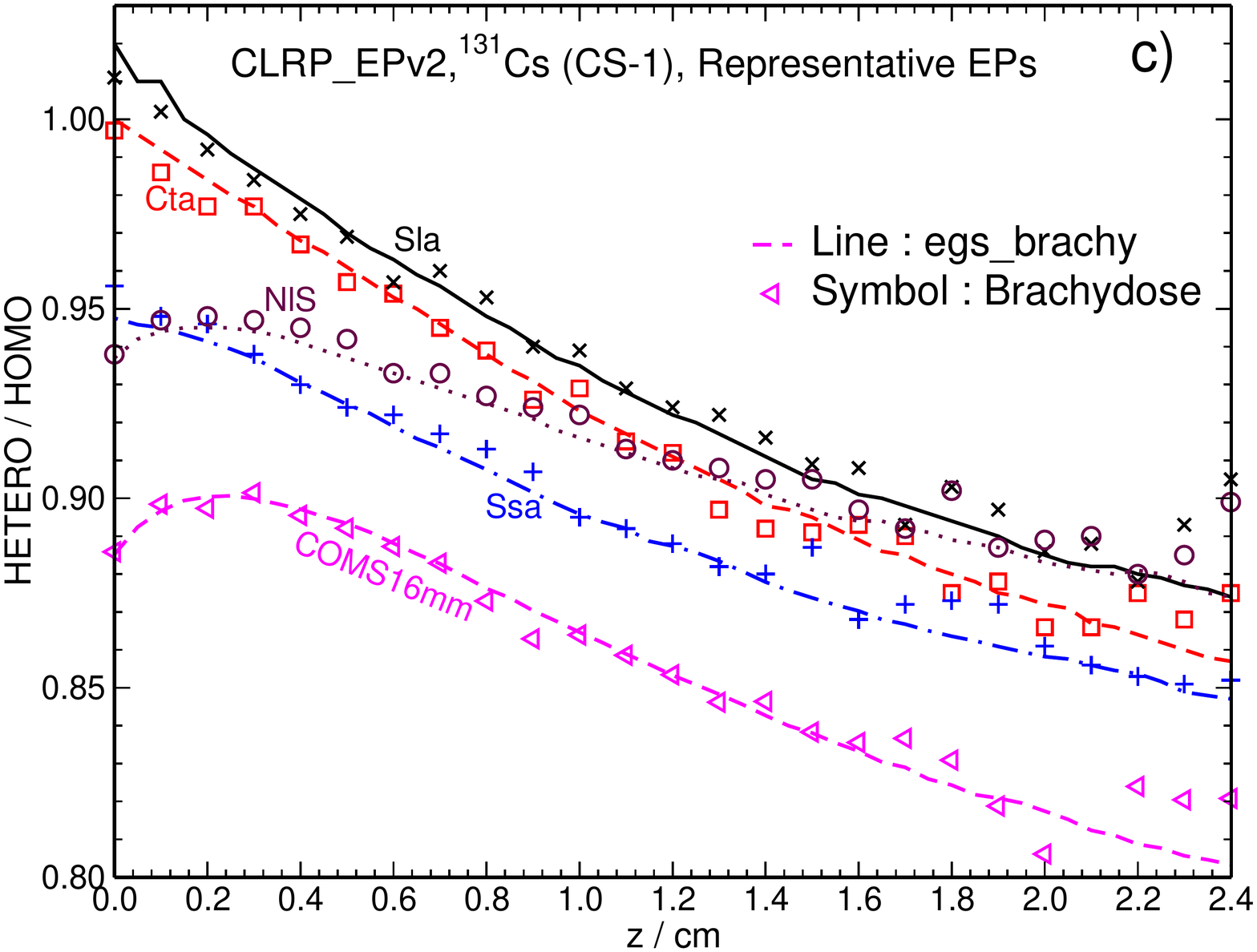}

\captionl{Comparison of calculated HETERO/HOMO doses along the central axis for \eb (lines) and \bd \cite{Le14a,tg221all} (symbols) for fully-loaded representative plaques (Cta: COMS-thin acrylic; Sla: Short lip-acrylic; Ssa: Stainless steel-acrylic; NlS: No lip-Silastic) with 16~mm COMS and BEBIG results shown for comparison for (a) \pd (model 200), (b) \io (model 6711; I25.S16 for BEBIG only), and (c) \cs (CS-1 Rev2) seeds.  Statistical uncertainties on the HETERO/HOMO dose ratio are at most 0.2\% for \eb and $1\%$ for \bdc  \cite{Le14a}.
 \label{Representative Eps_HETERO/HOMO_eb&BD} }
\end{center}
\end{figure}

%%%%%%%%%%%%%%%%%%%%%%%%
Figure~\ref{Representative Eps_HETERO/HOMO_eb&BD}\mfig{Representative Eps_HETERO/HOMO_eb&BD} provides HETERO/HOMO dose ratios  along the central axis for the representative plaques, comparing results to those from \bd from previous work \cite{tg221all, Le14a}, as well as 16~mm COMS and BEBIG results for comparison.  The \eb and \bd results are in good agreement, with percent differences of 0.1\% to 1.1\% for $z\leq 0.5$~cm (assumed tumour apex) and between 0.1\% to 4.5\% beyond tumor apex to the opposite side of sclera (where statistical uncertainties are larger).  The smallest deviations overall are observed for the ``Stainless steel - acrylic'' (Ssa) plaque, with average deviations of 0.2\% for $z\le0.5$~cm; conversely, the largest discrepancies are observed for ``No-lip - Silastic'' (NlS) with average deviations of 0.5\% for $z\le0.5$~cm and 0.9\% for larger $z$.  The \bd results for the HETERO/HOMO dose ratio are generally observed to fluctuate about the smoother lines presented by the \eb results, in accord with the larger statistical uncertainties on the \bd results (sub-1\% for $1\sigma$) in comparison with \eb (sub-0.2\%).  Apart from differences in the number of histories used to generate results for \eb and \bdc, there are also differences in the spectra of initialized particles, small differences in plaque dimensions (due to rounding in intermediate calculations to specify plaque geometries), and differences in the voxel scoring grid, as well as differences associated with the different codes as seen previously \cite{Th18,Ri11}.

Near the plaque (and sclera), the ``Short lip - acrylic'' plaque is observed to increase doses relative to HOMO doses, in accord with Lesp\'erance {\it et al} \cite{Le14a} who attributed this to the combined effects of fluorescence photons from the plaque and less photon attenuation through the (short) acrylic insert (for HETERO simulations) in comparison with water (HOMO).  For all other representative plaques, the HETERO/HOMO dose ratio  is $\le1$ near the plaque due to the lack of scatter and attenuation due to the plaque backing and insert.  For all plaque models (and seed types), the dominant effect of the plaque is a reduction in backscattering from regions behind it.  Scattered radiation accounts for more of the total dose in regions further from the plaque and so a reduction in the number of scattered photons when the plaque is present (in comparison with not, HOMO) causes decreasing HETERO/HOMO dose ratios with distance from the plaque \cite{Le14a}.

%%%%%%%%%%%%% HET/HOMO Representative models-egs_brachy& BrachyDose%%%%%%%%%

\FloatBarrier 
\FloatBarrier

%%%%%%%%%%%%%%%%%%%%%%%%%%%%%%%%%%%%%%%%
%%%%%%%%%%%%%%%%%%%%%%%%%%%%%%%%%%%%%%%%
\section{Data Format and Access}\label{data_acess}

The \E website is hosted at Carleton University, Ottawa, Ontario, Canada. The database is available online at \url{https://physics.carleton.ca/clrp/eye_plaque_v2}.
The main page of the database lists the 17 plaques which for the online datasets are available,  as well as details about plaque and seed models, MC simulation scenarios and normalization, and a spreadsheet of HOMO and HETERO doses along the central axis with 0.05~cm resolution.  The \db database includes results from more than 100 independent \eb simulations. For each plaque model, the following information is available on the \E database:

\begin{itemize}  
\item A to-scale image showing the eye plaque cross-section fully loaded with seeds in each of the Cartesian planes intersecting the origin, created using egs\_view images of actual egs++ model of the plaque and seeds.

\item A description of the plaque model, and seed coordinates according to publications and manufacturer information, as implemented in egs++. These models will be released for use with the open-source \eb application.

\item Files containing 3D dose distributions (.3ddose) in a zipped format for HOMO and HETERO scenarios, as well as HETsi for BEBIG plaques.  Doses are provided in units of dose rate per unit seed air kerma (Gy~h${}^{-1}$~U${}^{-1}$). Total dose (in Gy) in each voxel may be obtained by multiplying the data by the air kerma per seed (in units of U; 1~U $=$ 1 cGy cm$^2$ h$^{-1}$), and integrated over the treatment time, taking the exponential decay of the sources into account.  Statistical uncertainties (1 sigma) on the dose in each voxel is also included in the 3ddose file (as fractions of the local dose).
\item A figure for the depth dose curve along the plaque central axis  compared to the available \BD\cite{Th08,TR10,tg129all,tg221all,Le14a} and \MCNP\cite{MR08,Ri11} data for  HOMO and HETERO scenarios in units of dose/air kerma (Gy~h${}^{-1}$~U${}^{-1}$) for different seeds (where data are available).

\item A figure comparing $x,\,y$ transverse dose profiles at different depths ($z=0.5,\, 1$~cm) for HOMO and HETERO doses with the available corresponding values from \e simulated with \BD\cite{Th08,TR10,tg129all,tg221all,Le14a} (where data are available). For BEBIG plaques, data for HETsi scenario are presented.

\item For the 16~mm COMS plaque, a figure that summarizes percent dose differences of \eb with \bd \cite{Th08} or \mcnp \cite{MR08} along the central axis and at some organs at risk (\eg fovea, optic disk, lens, and lacrimal gland) \cite{Ri11}.

\item A figure comparing the ratio of HETERO/HOMO doses along the central axis with corresponding \BD\cite{tg221all,Le14a} data for each Representative plaques (\eg Cta, Sla, Ssa, and NlS) and different nuclide (\eg \pd, \io, \cs).

\end{itemize} 

%%%%%%%%%%%%%%%%%%%%%%%%%%%
%%%%%%%%%%%%%%%%%%%%%%%%%%%
\section{Potential Impact}\label{potential_impact}

The most recent AAPM task group reports pertaining to eye plaque brachytherapy, namely TG-129 \cite{tg129all} and TG-221 \cite{tg221all}, recommend that HETERO doses be calculated (or estimated) and reported in parallel with the traditional water-based (HOMO) TG-43 doses.  These recommendations are consistent with the recommendations of AAPM-ESTRO-ABG TG-186, which recommends adoption of model-based dose calculations where possible and reporting of their doses alongside TG-43.  Thus, the publicly-available 3D dose distributions and associated data provided in \db for each plaque and seed supports TG-129, 221, and 186 recommendations, enabling HETERO dose estimations and MC dose evaluations.  Furthermore, the plaque models developed herein will be distributed freely with \ebc, as will the seed models that are documented within the CLRP\_TG43v2 \cite{Sa20} database (\url{https://physics.carleton.ca/clrp/egs_brachy/seed_database_v2}).  The models may be used to carry out custom calculations within user-specified phantoms, including virtual patient models derived from patient (CT) datasets.  In addition, the eye plaque models may then be modified by users to develop models representative of their own practice.  Overall, the dose distributions and freely-distributed \eb plaque models support state-of-the art, advanced dose evaluations for ocular brachytherapy. 

%%%%%%%%%%%%%%%%%%%%%%%%%%%%%%%%%%
%%%%%%%%%%%%%%%%%%%%%%%%%%%%%%%%%
\section{Conclusion}\label{conclusion}

The  \E database offers accurate 3D dose distributions for more plaque models and radionuclides, plus lower statistical uncertainties than CLRP\_EPv1.  The \db database contains new datasets for 17 plaques (8 COMS, 5 BEBIG, and 4 representative plaques) and 3 radionuclides [\pac, \io (2 seed models), and \cscc], including both HOMO and HETERO (also HETsi for BEBIG) scenarios.  
The \db data are validated by comparison with \bd and \mcnp published data, for the plaque models for which such data exist: COMS plaque data is in good accord with previous \bd and \mcnp results, and representative plaques agree with previous \bd results.  For example, on the plaque central axis HETERO doses agree within 2\% for \bd and 5\% for \mcnp for COMS plaques, while doses agree within 4.5\% with \bd for the representative plaques.  The BEBIG plaques, modelled for the first time, differ only in media (elemental composition, density) from the COMS plaques, and differences in the relative HETERO/HOMO dose ratios may be understood on the basis of the higher-atomic number content of the BEBIG plaque backings in comparison with COMS, the lower density of the seed-carrier insert, as well as the differences in emitted photon spectra [\ioc: I125.S16 for BEBIG, 6711 for COMS].  The HETsi dose distributions offer the potential for users to tally dose distributions with different weights for custom-loading plaques with seeds of different activities.   The \db database provides reference 3D dose distributions that support the recommendations of AAPM TG-129 and TG-221.  The reference 3D dose distributions and benchmarked egs\_brachy (MC)  models  of  plaques and seeds   will  be  freely  distributed at (\url{https://physics.carleton.ca/clrp/eye_plaque_v2}), enabling advances in ocular brachytherapy research, dosimetry, and clinical practice.

%%%%%%%%%%%%%%%%%%%%%%%%%%%
%%%%%%%%%%%%%%%%%%%%%%%%%%%
\section{Acknowledgements}\label{ACK}

Chris Melhus and Mark Rivard are thanked for providing their MCNP5 data directly to facilitate comparisons.  The authors acknowledge  the  Natural  Sciences  and  Engineering  Research  Council  of Canada (NSERC), the Canada Research Chairs program, the Ministry of Research and Innovation of Ontario, and a Compute Canada National Resource Allocation.  

\section{Conflict of interest statement}
This work was partially supported by Eckert \& Ziegler BEBIG GmbH of Berlin, Germany.

%%%%%%%%%%%%%%%%%%%%%%%%%%%%%%%%%%%
%%%%%%%%%%%%%%%%%%%%%%%%%%%%%%%%%%%
%%%%%%%%%%%%%%%%%%%%%%%%%%%%%%%%%%%

\newpage

\section*{References}
%\addcontentsline{toc}{section}{\numberline{}References}
\vspace{-2.5cm}
\setlength{\baselineskip}{0.43cm}	%use for drafting
%
%
%
%%following must be set to point to local clrp.bib file
%%\bibliography{/home/drogers/bib/clrp}
%
%\bibliography{clrp}

\begin{thebibliography}{10}

\bibitem{tg221all}
R.~M. Thomson, K.~M. Furutani, T.~W. Kaulich, F.~Mourtada, M.~J. Rivard, C.~G.
  Soares, F.~M. Vanneste, and C.~S. Melhus,
\newblock AAPM recommendations on medical physics practices for ocular plaque
  brachytherapy: Report of task group 221,
\newblock Med. Phys. {\bf 47}, e92 -- e124 (2020).

\bibitem{tg129all}
S.-T. Chiu-Tsao, M.~A. Astrahan, P.~T. Finger, D.~S. Followill, A.~S. Meigooni,
  C.~S. Melhus, F.~Mourtada, M.~E. Napolitano, R.~Nath, M.~J. Rivard, D.~W.~O.
  Rogers, and R.~M. Thomson,
\newblock {Dosimetry of $^{125}$I and $^{103}$Pd COMS eye plaques for
  intraocular tumors: Report of Task Group 129 by the AAPM and ABS},
\newblock Med. Phys. {\bf 39}, 6161 -- 6184 (2012).

\bibitem{ABS-OOTF}
{ABS.~OOTF.~Committee},
\newblock {The American Brachytherapy Society consensus guidelines for plaque
  brachytherapy of uveal melanoma and retinoblastoma},
\newblock Brachytherapy {\bf 13}, 1 -- 14 (2014).

\bibitem{COMS18}
{COMS Group},
\newblock {The COMS randomized trial of iodine 125 brachytherapy for choroidal
  melanom. III. Initial Mortality findings. COMS report no. 18},
\newblock Arch. Ophthalmol. {\bf 105}, 969--982 (2001).

\bibitem{COMS28}
{COMS Group},
\newblock {The COMS randomized trial of iodine 125 brachytherapy for choroidal
  melanom. V. Twelve-year mortality rates and prognostic factors: COMS report
  no. 28},
\newblock Arch. Ophthalmol. {\bf 124}, 1684--1693 (2006).

\bibitem{Li17}
A.~J. Lin, Y.~J. Rao, S.~Acharya, J.~Schwarz, P.~K. Rao, and P.~Grigsby,
\newblock Patterns of care and outcomes of proton and eye plaque brachytherapy
  for uveal melanoma: review of the National Cancer Database,
\newblock Brachytherapy {\bf 16}, 1225--1231 (2017).

\bibitem{Ri17}
{M. J. Rivard {\em et al.}},
\newblock {Supplement 2 for the 2004 update of the AAPM Task Group No. 43
  Report: Joint recommendations by the AAPM and GEC-ESTRO},
\newblock Med. Phys. {\bf 44}, e297 -- e338 (2017).

\bibitem{Ri11}
{M. J. Rivard \em et.al.},
\newblock {Comparison of dose calculation methods for brachytherapy of
  intraocular tumors},
\newblock Med. Phys. {\bf 38}, 306 -- 316 (2011).

\bibitem{MR08}
C.~S. Melhus and M.~J. Rivard,
\newblock {COMS eye plaque brachytherapy dosimetry simulations for
  ${}^{103}$Pd, ${}^{125}$I, and ${}^{131}$Cs},
\newblock Med. Phys. {\bf 35}, 3364 -- 3371 (2008).

\bibitem{Th08}
R.~M. Thomson, R.~E.~P. Taylor, and D.~W.~O. Rogers,
\newblock {Monte Carlo dosimetry for $^{125}$I and $^{103}$Pd eye plaque
  brachytherapy},
\newblock Med. Phys. {\bf 35}, 5530 -- 5543 (2008).

\bibitem{TR10}
R.~M. Thomson and D.~W.~O. Rogers,
\newblock {Monte Carlo dosimetry for $^{125}$I and $^{103}$Pd eye plaque
  brachytherapy with various seed models},
\newblock Med. Phys. {\bf 37}, 368 -- 376 (2010).

\bibitem{Th10}
R.~M. Thomson, K.~M. Furutani, J.~S. Pulido, S.~L. Stafford, and D.~W.~O.
  Rogers,
\newblock {Modified COMS plaques for $^{125}$I and $^{103}$Pd iris melanoma
  brachytherapy},
\newblock Int. J. Radiat. Oncol. Biol. Phys. {\bf 78}, 1261 -- 1269 (2010).

\bibitem{Le14}
M.~Lesperance, M.~{Inglis-Whalen}, and R.~M. Thomson,
\newblock {Model-based dose calculations for COMS eye plaque brachytherapy
  using an anatomically realistic eye phantom},
\newblock Med. Phys. {\bf 41}, 021717 (12pp) (2014).

\bibitem{Le14a}
M.~Lesperance, M.~Martinov, and R.~M. Thomson,
\newblock {Monte Carlo dosimetry for $^{103}$Pd, $^{125}$I, and $^{131}$Cs
  ocular brachytherapy with various plaque models using an eye phantom},
\newblock Med. Phys. {\bf 41}, 031706 (14pp) (2014).

\bibitem{tg186all}
L.~Beaulieu, A.~C. Tedgren, J.-F. Carrier, S.~D. Davis, F.~Mourtada, M.~J.
  Rivard, R.~M. Thomson, F.~Verhaegen, T.~A. Wareing, and J.~F. Williamson,
\newblock {Report of the Task Group 186 on model-based dose calculation methods
  in brachytherapy beyond the TG-43 formalism: Current status and
  recommendations for clinical implementation},
\newblock Med. Phys. {\bf 39}, 6208 -- 6236 (2012).

\bibitem{Ch16}
M.~Chamberland, R.~E.~P. Taylor, D.~W.~O. Rogers, and R.~M. Thomson,
\newblock {egs\_brachy: a versatile and fast Monte Carlo code for
  brachytherapy},
\newblock Phys. Med. Biol. {\bf 61}, 8214 -- 8231 (2016).

\bibitem{Th18}
R.~M. Thomson, R.~E.~P. Taylor, M.~J.~P. Chamberland, and D.~W.~O. Rogers,
\newblock Reply to Comment on 'egs\_brachy: a versatile and fast Monte Carlo
  code for brachytherapy',
\newblock Phys. Med. Biol. {\bf 63}, 038002(5pp) (2018).

\bibitem{Sa20}
H.~Safigholi, M.~J.~P. Chamberland, R.~E.~P. Taylor, C.~H. Allen, M.~P.
  Martinov, D.~W.~O. Rogers, and R.~M. Thomson,
\newblock {Update of the CLRP TG-43 parameter database for low-energy
  brachytherapy sources},
\newblock Med. Phys. {\bf 47}, 4656--4669 (2020).

\bibitem{Ma20a}
E.~Mainegra-Hing, D.~Rogers, F.~Tessier, and B.~Walter,
\newblock {The EGSnrc Code System: Monte Carlo Simulation of Electron and
  Photon Transport,NRCC Report PIRS-701},
\newblock Technical report, National Research Council Canada, Ottawa, Canada.
  https://nrc-cnrc.github.io/EGSnrc/, (2020).

\bibitem{BH87}
M.~J. Berger and J.~H. Hubbell,
\newblock { XCOM: Photon cross sections on a personal computer},
\newblock Report NBSIR87--3597, National Institute of Standards Technology
  (NIST), Gaithersburg, MD~20899, U.S.A., 1987.

\bibitem{Ma20}
E.~Mainegra-Hing, D.~W.~O. Rogers, R.~Townson, B.~R.~B. Walters, F.~Tessier,
  and I.~Kawrakow,
\newblock {The EGSnrc g application Technical Report PIRS-3100},
\newblock Technical report, National Research Council Canada, Ottawa, Canada.
  https://nrc-cnrc.github.io/EGSnrc/, (2020).

\bibitem{ICRU90a}
ICRU,
\newblock {Report 90: Key data for ionizing-radiation dosimetry: measurement
  standards and applications},
\newblock J ICRU {\bf 14}, 1--110 (2014).

\bibitem{Cu15a}
S.~E.~M. Cutsinger, K.~M. Furutani, R.~M. Forsman, and S.~M. Corner,
\newblock {Seed coordinates of a new COMS‐like 24 mm plaque verified using
  the FARO Edge},
\newblock J of App Clin Med Phys {\bf 16}, 293 -- 301 (2015).

\bibitem{Po03}
D.~P. Potter,
\newblock {Treatment of intraocular melanoma: new concepts},
\newblock Bull Mem Acad R Med Belg {\bf 158}, 103–111 (2003).

\bibitem{Fi99b}
P.~T. Finger, A.~Berson, and A.~Szechter,
\newblock {Palladium-103 plaque radiotherapy for choroidal melanoma},
\newblock Ophthalmology {\bf 106}, 606 -- 613 (1999).

\bibitem{Pu04}
I.~Puusaari, J.~Heikkonen, and T.~Kivela,
\newblock {Effect of radiation dose on ocular complications after iodine
  brachytherapy for large uveal melanoma: empirical data and simulation of
  collimating plaques},
\newblock Invest Ophthalmol Vis Sci {\bf 45}, 3425–3434 (2004).

\bibitem{Gr04}
D.~Granero, J.~P{\'e}rez-Calatayud, F.~Ballester, E.~Casal, and J.~De~Frutos,
\newblock Dosimetric study of the ROPES eye plaque,
\newblock Med. Phys. {\bf 31}, 3330--3336 (2004).

\bibitem{MW02}
J.~I. Monroe and J.~F. Williamson,
\newblock { Monte Carlo-aided dosimetry of the Theragenics TheraSeed Model 200
  $^{103}$Pd interstitial brachytherapy seed},
\newblock Med. Phys. {\bf 29}, 609 -- 621 (2002).

\bibitem{Do06}
J.~Dolan, Z.~Li, and J.~F. Williamson,
\newblock {Monte Carlo and experimental dosimetry of an $^{125}$I brachytherapy
  seed},
\newblock Med. Phys. {\bf 33}, 4675 -- 4684 (2006).

\bibitem{Ri07}
M.~J. Rivard,
\newblock {Brachytherapy dosimetry parameters calculated for a $^{131}$Cs
  source},
\newblock Med. Phys. {\bf 34}, 754 -- 762 (2007).

\bibitem{He00}
H.~Hedtj\"arn, G.~A. Carlsson, and J.~F. Williamson,
\newblock {Monte Carlo-aided dosimetry of the symmetra model I25.S06 I$^{125}$,
  interstitial brachytherapy seed},
\newblock Med. Phys. {\bf 27}, 1076--1085 (2000).

\bibitem{NNDC}
{Brookhaven National Laboratory, National Nuclear Data Center},
\newblock {http://www.nndc.bnl.gov/nudat2}.

\bibitem{NCRP58}
{NCRP Report 58},
\newblock { A Handbook of Radioactivity Measurements Procedures},
\newblock NCRP Publications, 7910 Woodmont Avenue, Bethesda, MD. 20814 USA
  (1985).

\end{thebibliography}
%
%%\bibliographystyle{/home/drogers/tex/bst/medphy}
%\bibliographystyle{medphy}

\end{document}